\documentclass{aa} 
\usepackage{lineno}
\usepackage{float}
\usepackage{graphicx}
\usepackage{caption}
\usepackage{subcaption}
\usepackage{tabu}
\usepackage{xcolor}
\usepackage{graphicx}
\usepackage{txfonts}
\usepackage[breaklinks=true]{hyperref}
\usepackage{graphicx,colortbl}
\usepackage{multirow}

\usepackage{natbib,twoopt}

\bibpunct{(}{)}{;}{a}{}{,} 
\makeatletter
  \newcommandtwoopt{\citeads}[3][][]{\href{http://adsabs.harvard.edu/abs/#3}%
    {\def\hyper@linkstart##1##2{}%
     \let\hyper@linkend\@empty\citealp[#1][#2]{#3}}}
  \newcommandtwoopt{\citepads}[3][][]{\href{http://adsabs.harvard.edu/abs/#3}%
    {\def\hyper@linkstart##1##2{}%
     \let\hyper@linkend\@empty\citep[#1][#2]{#3}}}
  \newcommandtwoopt{\citetads}[3][][]{\href{http://adsabs.harvard.edu/abs/#3}%
    {\def\hyper@linkstart##1##2{}%
     \let\hyper@linkend\@empty\citet[#1][#2]{#3}}}
  \newcommandtwoopt{\citeyearads}[3][][]%
    {\href{http://adsabs.harvard.edu/abs/#3}
    {\def\hyper@linkstart##1##2{}%
     \let\hyper@linkend\@empty\citeyear[#1][#2]{#3}}}
\makeatother

\begin{document}

   \title{Normal and counter Evershed flows in the  photospheric penumbra of a sunspot}

   \subtitle{SPINOR 2D inversions of Hinode-SOT/
SP observations }

   \author{A. Siu-Tapia
          \inst{1}
          \and
	A. Lagg\inst{1} \and
          S. K. Solanki\inst{1,}\inst{2}  \and
          M. van Noort \inst{1} \and
	J. Jur\v{c}\'ak\inst{3}
          }

   \institute{Max-Planck-Institut f\"ur Sonnensystemforschung, Justus-von-Liebig-Weg 3,
37077 G\"ottingen, Germany.\\
              \email{[siu;solanki;lagg;vannoort]@mps.mpg.de}
         \and
             School of Space Research, Kyung Hee University, Yongin, 446-701 Gyeonggi, Republic of Korea.
	\and
	Astronomical Institute of the Academy of Sciences, Fri\v{c}ova 298, 25165 Ond\v{r}ejov, Czech Republic
	\\
              \email{jurcak@asu.cas.cz}
             }


 
  \abstract
   {The Evershed effect, a nearly horizontal outflow of material seen in the penumbrae of sunspots at the photospheric layers, is a common characteristic of well-developed penumbrae, but is still not well understood.  Even less is known about photospheric horizontal inflows in the penumbra, also known as counter Evershed flows.}
   {Here we present  a rare feature observed in the penumbra of the main sunspot of AR NOAA 10930. This spot displays the normal Evershed outflow in most of the penumbra, but  harbors a fast photospheric inflow of material over a large sector of the disk-center penumbra. We investigate the driving forces of both, the normal and the counter Evershed flows.
}
   {
We invert the spectropolarimetric data from Hinode SOT/SP using the SPINOR 2D inversion code, which allows us to derive  height-dependent maps of the relevant physical parameters in the sunspot. These maps show considerable fine structure.  Similarities and differences between the normal Evershed outflow and the counter Evershed flow  are investigated.  }
   {In both, the normal and the counter Evershed flows, the material flows from regions of weak ($\sim1.5$ kG to $ \sim2$ kG) to stronger fields. The sources and sinks of both penumbral flows display opposite field polarities; with the sinks (tails of filaments) harboring local enhancements in temperature, which are nonetheless colder than their sources (heads of filaments). 
 } 
   {
The anti-correlation of the gradients in the temperature and magnetic pressure between the endpoints of the filaments from the two distinct penumbral regions  is compatible with both the convective driver and the siphon flow scenarios. 
A geometrical scale of the parameters is necessary to determine which the dominant force driving the flows is. 
}

   \keywords{Sun: photosphere--
                sunspots --
                Sun: surface magnetism
               }

  \maketitle

\section{Introduction}

The penumbrae of sunspots are strongly magnetized media (with field strengths of $\sim$1 to 2 kG) where convection is expected to be almost completely suppressed according to simple estimates \citep{Biermann1941, Cowling1953, Meyer1974, Jahn1994}. Consequently, the penumbral brightness is expected to be  much lower than the observed one, which is $\sim$75-80 $\%$ that of the quiet Sun, integrated over wavelength. This fact points towards some level of convection taking place in the penumbra to account for its observed brightness.
However, how the energy is transported in the penumbra is still one of the major open questions in solar physics. Detailed reviews pointing out this open problem and providing discussions on some proposed solutions have been given by, e. g.,  \citet{Solanki2003,Thomas2004,Thomas2008,Borrero2009,Scharmer2009,Schlichenmaier2009,Tritschler2009,Bellot2010,Borrero2011,Rempel2011a}.

The filamentary structure of the penumbra might provide one of the main clues to gaining insight into this question. The penumbral magnetic field consists of two major components. The first are \textit{spines}, seen as relatively dark regions where the magnetic field is stronger and more vertical. The others are \textit{intraspines/filaments}, seen as elongated bright channels where the magnetic field is weaker and more horizontal \cite[see review by][] {Borrero2011}. 

 Various models have been proposed to account for the filamentary structure of the penumbra. One of these, the \textit{embedded flux tube}, is empirical in nature and was proposed by \citet{Solanki1993}. This model mainly describes the complex 3-D structure of the field to explain asymmetric Stokes $V$ profiles. According to it,  nearly horizontal magnetic flux tubes forming the intraspines are embedded in more vertical background magnetic fields (spines) in the penumbra. The downward pumping mechanism  \citep{Thomas2002} was proposed to explain the origin of field lines that return  into the solar surface at the outer penumbra. 
Another idea to account for the penumbral filaments is the \textit{field-free gap model} \citep{Choudhuri1986,Scharmer2006, Spruit2006}, where the penumbral bright filaments are described as regions where the vertical component of the magnetic field  vanishes as a result of the interaction with the non-magnetized gas that rises due to convection into a background with more oblique fields. 

These  models concentrate on the configuration of the magnetic field in the penumbra. However, the appearance of a penumbra is always associated with a distinctive gas flow, i.e., the Evershed flow \cite[EF;][]{Evershed1909} and, therefore, this must also be taken into account by these models. The EF is the most prominent dynamic phenomenon in sunspots: An outward directed flow observed in the photospheric layers of
penumbrae with speeds of several km s$^{-1}$ and large inclinations to the vertical. 
This phenomenon is thought to be closely related to the filamentary structure of the penumbra \cite[e. g.,][]{Borrero2011} and given its ubiquity, it is expected to play a central role in the energy transport in the penumbrae of sunspots.

The EF  is usually observed as a blueshift  of photospheric spectral lines in the disk-center-side part of the penumbra and a corresponding redshift  in the limb-side part of the penumbra. This is generally interpreted as a radial, nearly horizontal outflow of matter.
The EF is height dependent: in the photosphere the line shifts decrease rapidly with height of line formation  \citep{StJohn1913,Maltby1964,Borner1992}. Moreover, in the chromosphere the line shifts change sign (inverse EF), with the center-side part of the penumbra now showing redshifts \citep{StJohn1913,Borner1992,Tsiropoula2000}. This is taken to be the signature of an inflow of material. Most of the mass flowing outwards in the photosphere returns to the solar interior within the penumbra, in opposite polarity downflow channels \citep{Westendorp1997,Westendorp2001}, although a part of the Evershed flow continues in the canopy of the sunspot  \citep{Solanki1994}.

The origin and driving physical mechanisms of the EF have been subject of considerable controversy for decades.  Some models describe it as a siphon flow driven by a gas pressure difference between the footpoints of arched magnetic 
flux tubes \citep{Meyer1968,Thomas1992,Montesinos1997,Thomas2006}; while, in others, the EF is explained as a flow along magnetic flux tubes driven by a form of convection \citep{Jahn1994,Schlichenmaier1998,Schlichenmaier2003}. 
According to a more recent proposal by \citet{Scharmer2006},  the EF  takes place in field-free gaps below the penumbral field.

Recent 3-D MHD simulations for penumbral fine structure \citep{Heinemann2007,Rempel2009b,Rempel2009a,Rempel2011a,Rempel2011b, Rempel2012}  display a flow very similar to the EF.
In such simulations the EF has typically been interpreted to be a consequence of  overturning convection: the hot gas rising from below the surface is deflected by the inclined magnetic field of the penumbra, producing a fast flow toward the sunspot border. Part of the rising gas turns over laterally and dips down below the solar surface.  The convective cells are then elongated in the preferred direction imposed by the magnetic field (the radial direction), forming penumbral filaments with a fast Evershed outflow along their axes and weaker downflows towards their sides, much in the same way as in  quiet sun  granules.

The relative importance of various forces for driving the EF is still a matter of debate. 
The reason is that along a filament harboring a flow, both the temperature and the magnetic field show the correct sign of the gradient along the filament \citep{Tiwari2013}.

In this paper, we report and 
study the characteristics  of an atypical  photospheric inflow observed by Hinode  SOT/SP over a considerable sector of the penumbra in the main sunspot of NOAA AR 10930. 
Observations of photospheric counter Evershed flows have been reported rather rarely, and are usually restricted to very narrow channels  \cite[see, e. g.,][]{Kleint2013,Louis2014}, or are transient during the formation phase of the penumbra, \cite[see e. g.,][]{Schlichenmaier2012,Romano2014,Murabito2016}.  
The observation of such a large inflowing region in a fully developed penumbra is to our knowledge unique.
 By comparing the properties of the anomalous counter EF with the well-known photospheric Evershed outflow, we hope to learn more about the drivers of both flows.

  This work is organized as follows: In section 2
we describe our data and inversion technique. In section 3, results are presented and  are discussed in section 4.  Finally,  in section 5 we draw our conclusions.

\section{Observational data and analysis techniques}
\subsection{Observations}

\begin{figure*}[htb]
    \centering
 \begin{subfigure}[width=17cm]{\textwidth}
    \begin{subfigure}[]{0.5\textwidth}
        \centering
        \includegraphics[width=\textwidth]{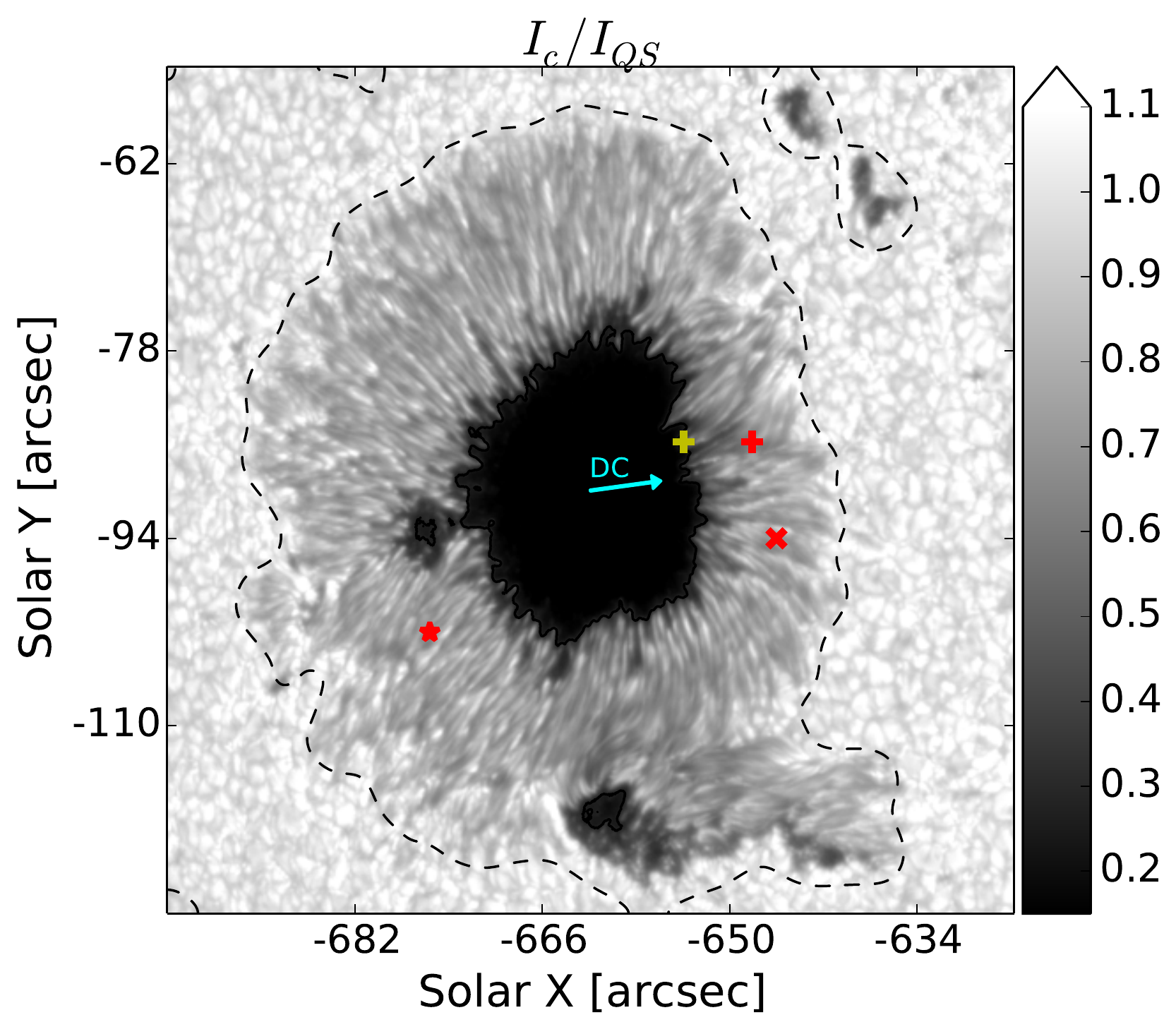}
        \caption{}
	\label{fig:1a}
    \end{subfigure}%
    ~ 
    \begin{subfigure}[width=17cm]{0.5\textwidth}
        \centering
        \includegraphics[width=\textwidth]{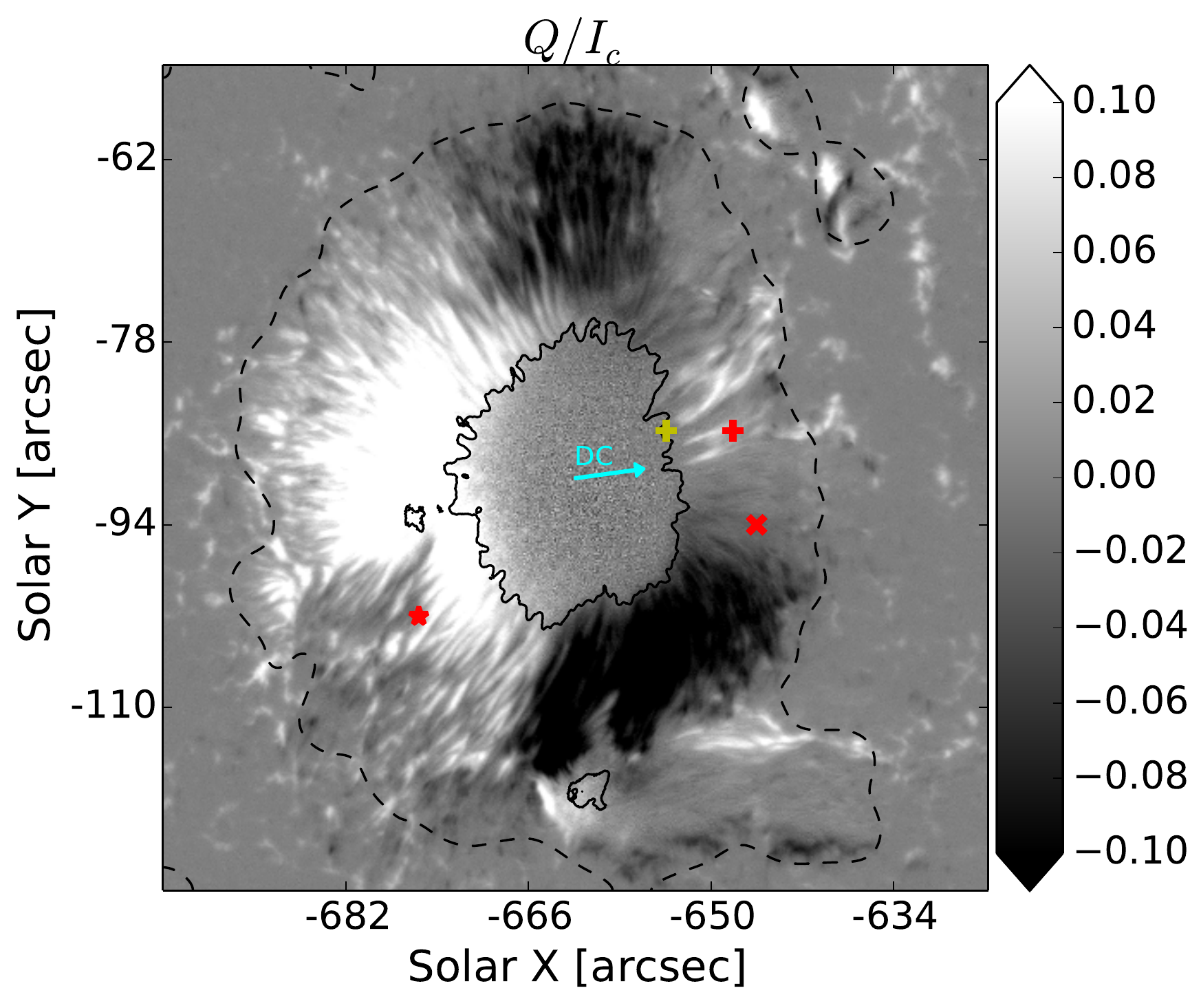}
        \caption{}
	\label{fig:1b}
    \end{subfigure}
      \end{subfigure}

\centering
 \begin{subfigure}[width=17cm]{\textwidth}
    \begin{subfigure}[]{0.5\textwidth}
        \centering
        \includegraphics[width=\textwidth]{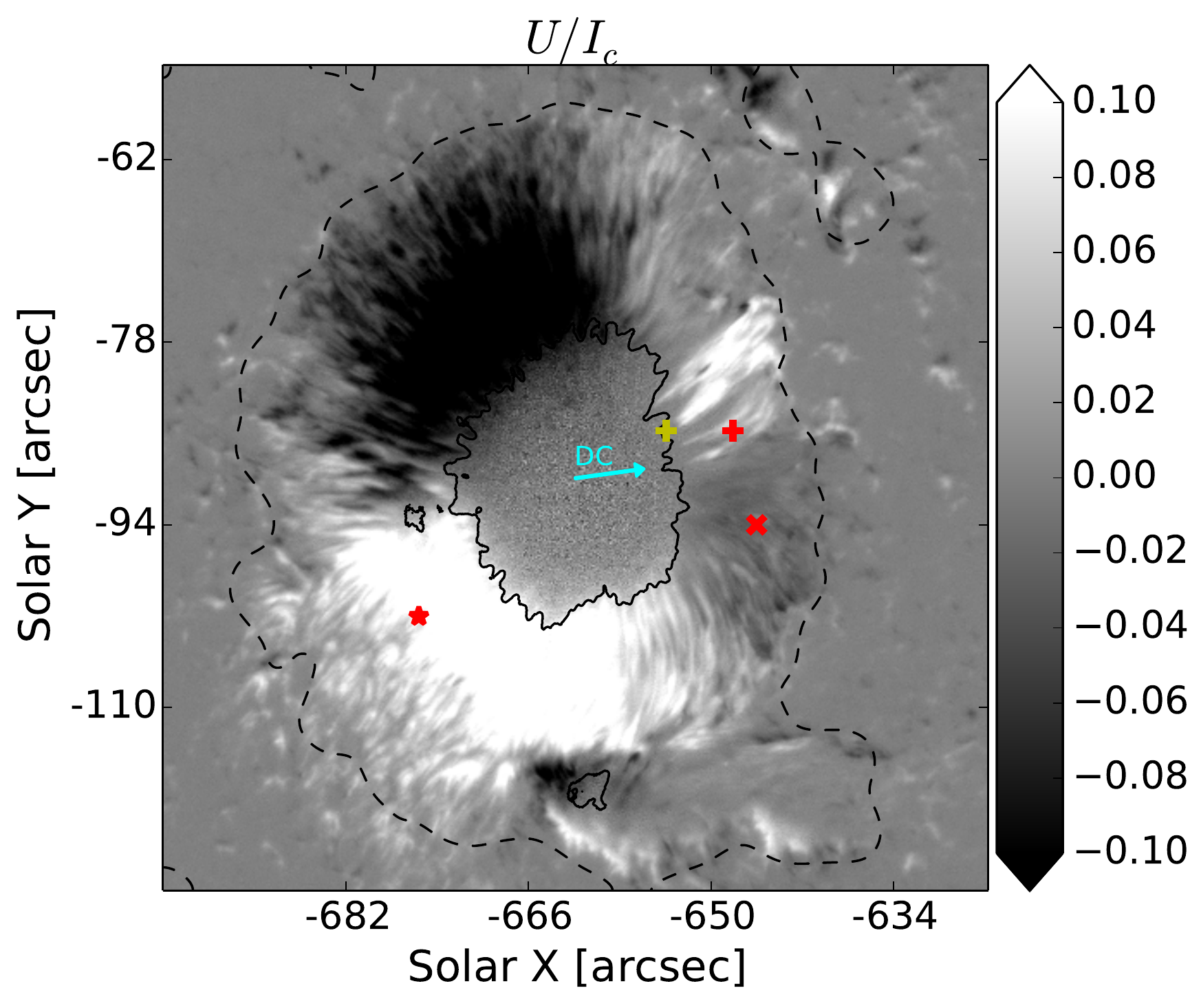}
        \caption{}
	\label{fig:1c}
    \end{subfigure}%
    ~ 
    \begin{subfigure}[width=17cm]{0.5\textwidth}
        \centering
        \includegraphics[width=\textwidth]{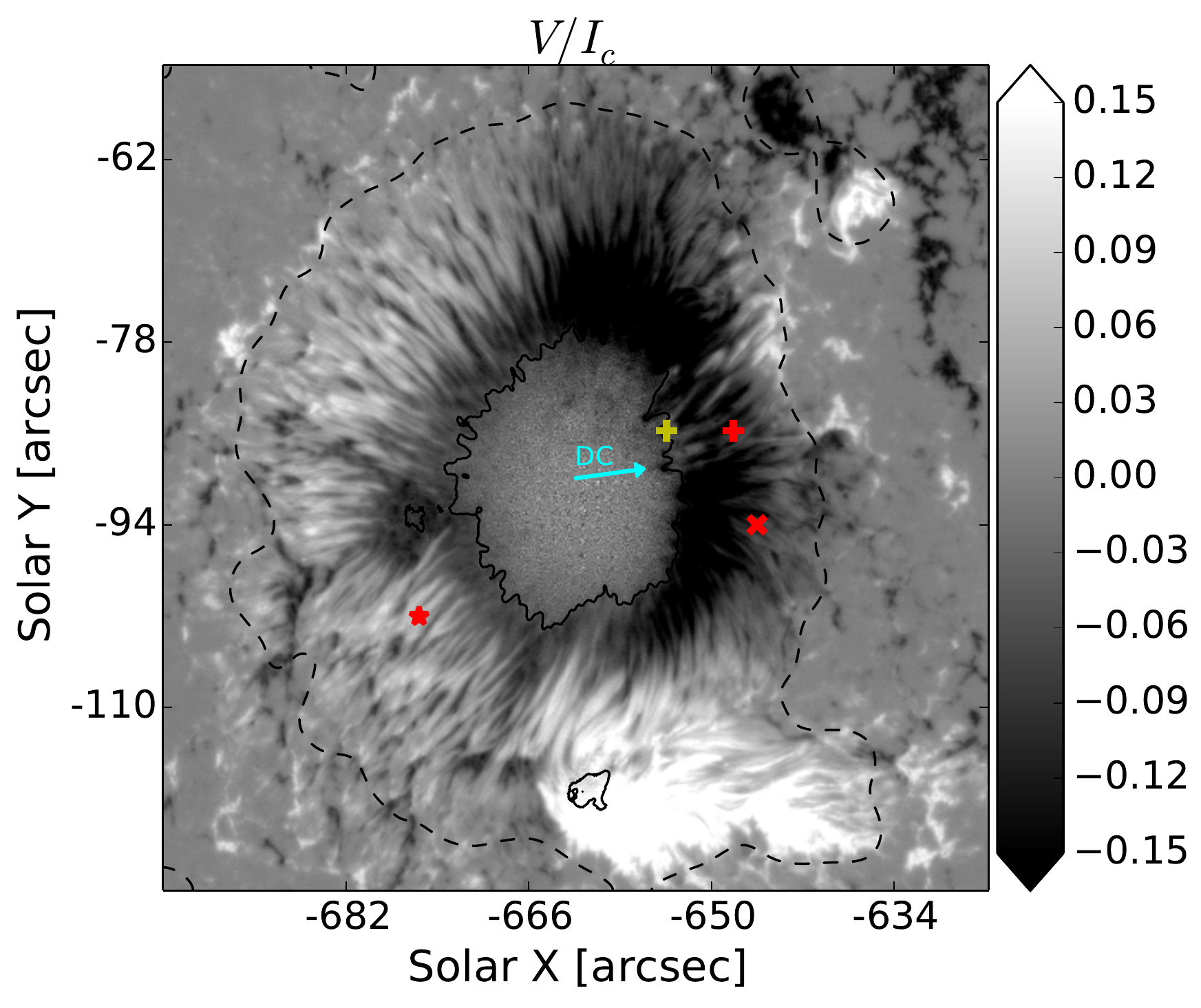}
        \caption{}
	\label{fig:1d}
    \end{subfigure}
      \end{subfigure}

     \caption{Stokes maps observed by the Hinode SOT/SP in the main sunspot of the  NOAA AR 10930 on December 08, 2006.  Panel (a) shows the continuum intensity $I_{c}$ normalized to the mean continuum value in the quiet sun, $I_{QS}$. Panels (b), (c) and (d) show the maps of the Stokes parameters $Q$,  $U$ and $V$, respectively, normalized to local $I_c$. The Stokes  $Q$,  $U$ and $V$ maps were constructed at $-0.1 \AA$ from the 6302.5 $\AA$ line core (the selected wavelength is indicated by vertical green lines in  Figure \ref{fig:1*}).
 The umbra-penumbra boundary (black solid contour) was placed at $I_{c}/I_{QS}=0.26$ and the external penumbral boundary (black dashed contour) is at $I_{c}/I_{QS}=0.94$.  The cyan arrows point towards the disk center. 
Red markers show three selected pixels, one located in  the limb-side penumbra (`*'), and two located in the center-side penumbra (`x' and `+', respectively); their corresponding Stokes profiles are shown in Figures \ref{fig:1a*}, \ref{fig:1b*} and \ref{fig:1c*}, respectively. The yellow cross shows a pixel close to the inner penumbral boundary where the inversions give $B>7$ kG at $\log(\tau)=0$ and whose Stokes profiles are shown in Figure \ref{fig:1d*}
}
\label{fig:1}
\end{figure*}

For our study, we utilize spectropolarimetric observations in the Fe I 6301.5 and 6302.5 $\AA$
lines from the Hinode spectropolarimeter of the Solar Optical Telescope (SOT/SP)  \citep{Kosugi2007,Lites2007,Ichimoto2008,Lites2013b} of the main sunspot of the active region (AR) NOAA 10930 
on  December 08, 2006 (see Figure \ref{fig:1}).
The SOT/SP provided us with a full-Stokes dataset with a spatial sampling of
$\sim$ $0.16''$/pixel while operating in \textit{normal mode} (see \citet{Lites2013b} for a detailed description of the SP instrument).

The sunspot umbra displays a negative magnetic polarity. It
was observed at (S91$''$, W698$''$), i.e. at an heliocentric angle $\theta \approx 47^{\circ}$. The observations
were reduced with the corresponding routines of the Solar-Soft
package \citep{Lites2013}.

\begin{figure*}[htb]
    \centering
 \begin{subfigure}[width=17cm]{\textwidth}
    \begin{subfigure}[]{0.5\textwidth}
        \centering
        \includegraphics[width=\textwidth]{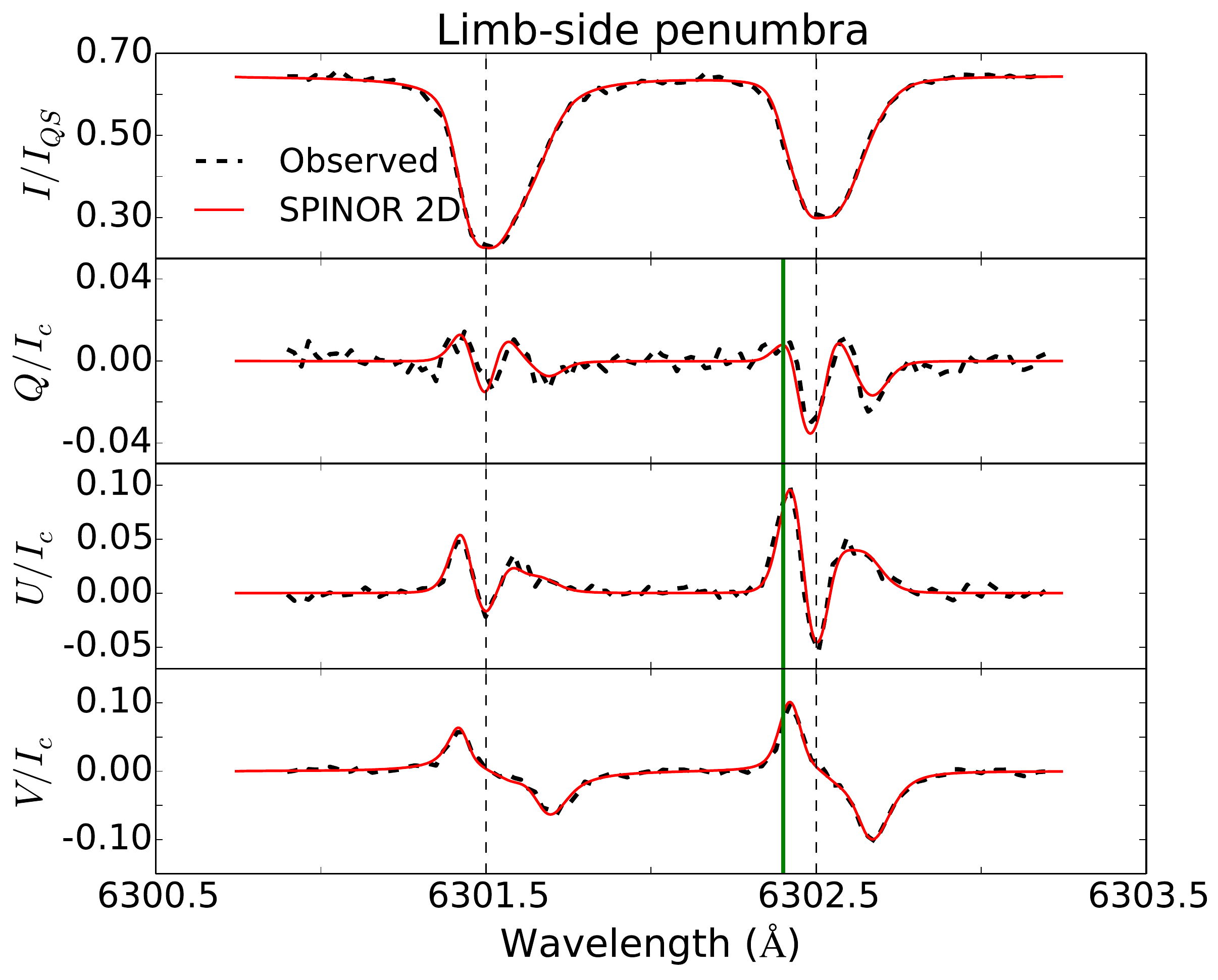}
        \caption{}
	\label{fig:1a*}
    \end{subfigure}%
    ~ 
    \begin{subfigure}[width=17cm]{0.5\textwidth}
        \centering
        \includegraphics[width=\textwidth]{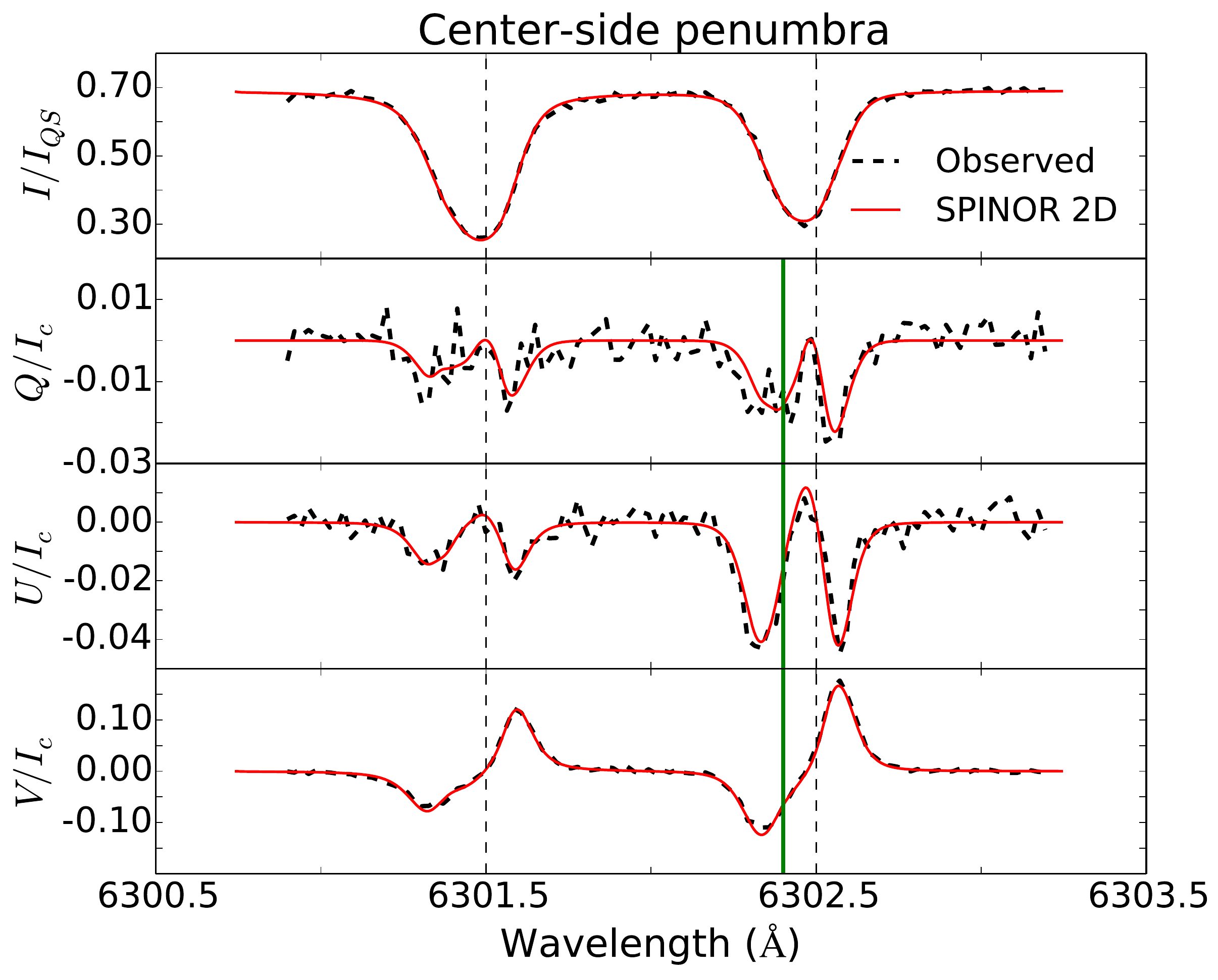}
        \caption{}
	\label{fig:1b*}
    \end{subfigure}
      \end{subfigure}

\centering
 \begin{subfigure}[width=17cm]{\textwidth}
    \begin{subfigure}[]{0.5\textwidth}
        \centering
        \includegraphics[width=\textwidth]{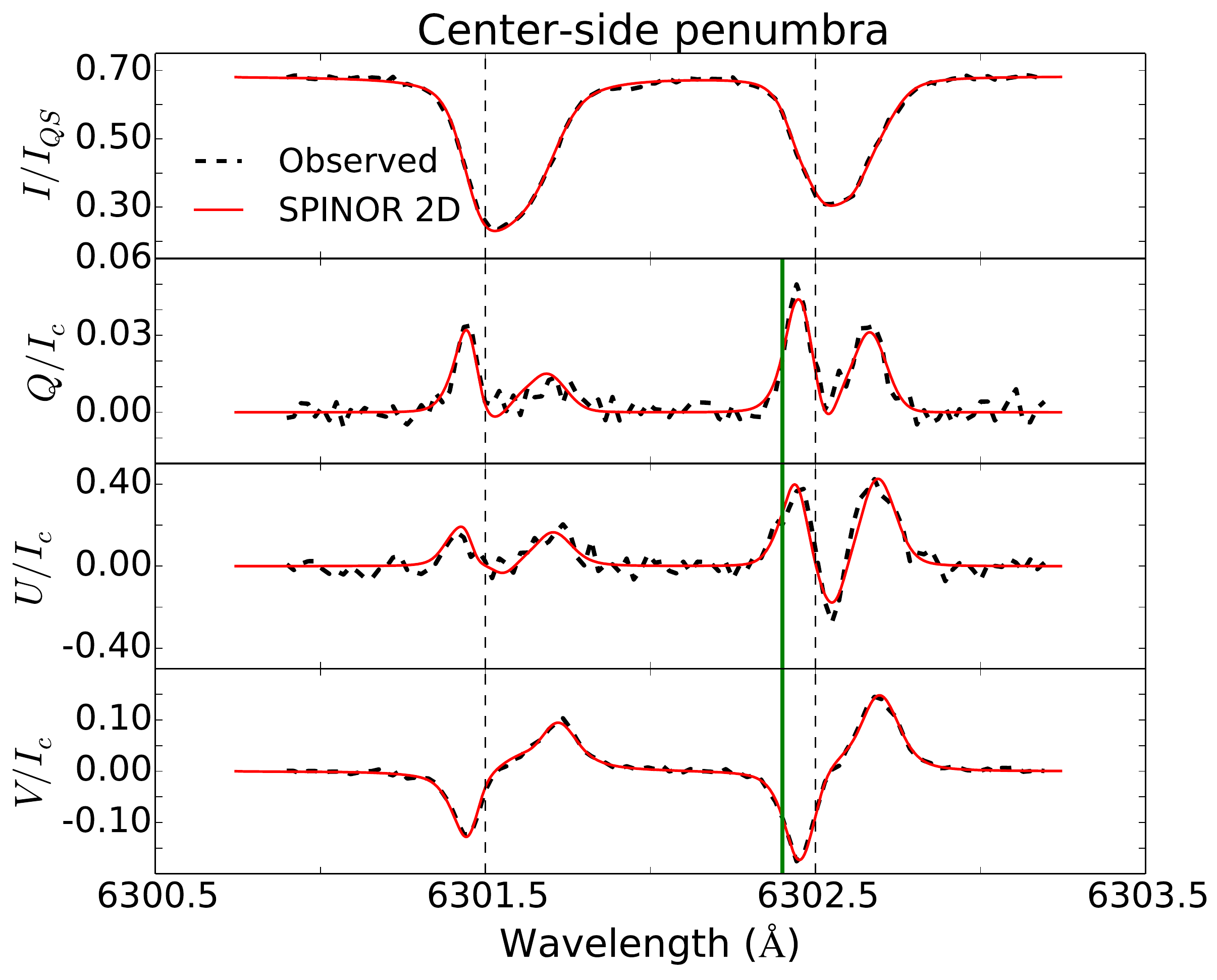}
        \caption{}
	\label{fig:1c*}
    \end{subfigure}%
    ~ 
    \begin{subfigure}[width=17cm]{0.5\textwidth}
        \centering
        \includegraphics[width=\textwidth]{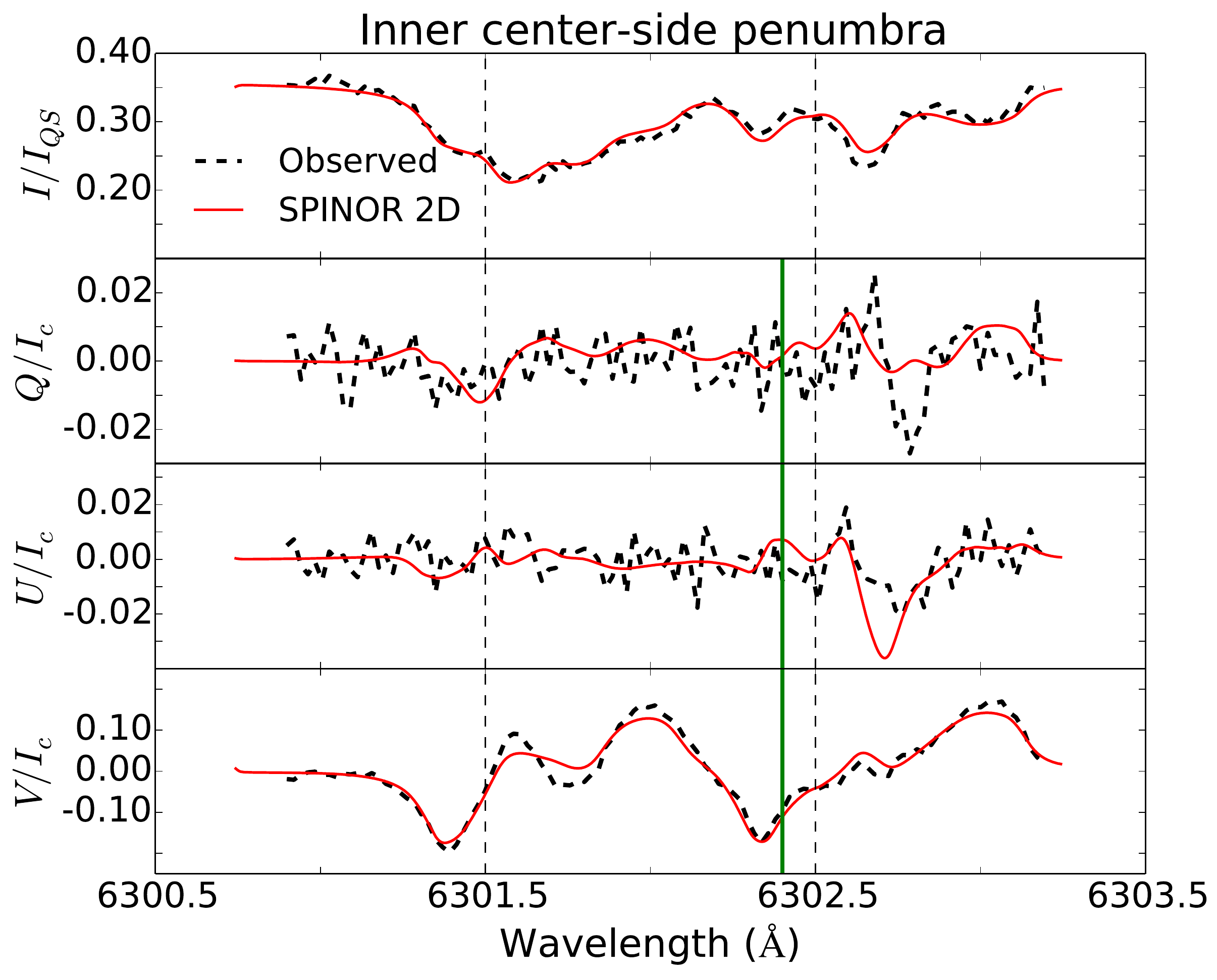}
        \caption{}
	\label{fig:1d*}
    \end{subfigure}
      \end{subfigure}

     \caption{Observed Stokes profiles (dashed lines) and SPINOR 2D best-fits (solid red lines), in the wavelength interval $6300.90-6303.19$ $\AA$, at the location of the red markers shown in Figure \ref{fig:1}: (a) From top to bottom: Stokes $I/I_{QS}$, $Q/I_c$, $U/I_c$ and $V/I_c$ in a  pixel from the limb-side penumbra (red `*' in Figure \ref{fig:1}); vertical green lines in the blue wing of 6302.5 $\AA$ indicate the selected wavelength  used to construct the Stokes $Q/I_c$, $U/I_c$ and $V/I_c$  maps in Figure \ref{fig:1} ($-0.1 \AA$ from 6302.5 $\AA$ line core). Vertical dashed lines were placed at $6301.5$ $\AA$ and at $6302.5$ $\AA$, respectively. (b), (c) and (d) show the Stokes profiles in three pixels from the center-side  penumbra (red `x',  red `+' and yellow `+'  in Figure \ref{fig:1}, respectively) in the same format as in (a). The SPINOR 2D best fits of profiles in panel (d) give $B\sim$ 6.4, 8.0 and 8.3 kG, $\gamma \sim$ 141, 148 and 145$^{\circ}$, and $v_{LOS}\sim$ 5.6, 8.3 and 9.3 km s$^{-1}$ at $\log(\tau)=-$2.0, $-$0.8 and 0, respectively, with $\chi^2 =$ 14.}
\label{fig:1*}
\end{figure*}

\subsection{Inversions}

We infer the atmospheric properties of the sunspot by inverting
the observational data with the
spatially coupled version \citep{vannoort2012} of the
SPINOR inversion code \citep{Frutiger2000} which 
uses the STOPRO routines \citep{Solanki1987} to
solve the radiative
transfer equations for polarized light  under the assumption of local thermodynamic equilibrium
(LTE).

The  inversion code calculates the best fit full-Stokes
spectra of an atmospheric model described by a selected number
of atmospheric parameters, specified at a number of optical
depth positions, and interpolated using a bicubic spline
approximation. 
The spatially coupled version is able to invert
the observed data while taking into account the spatial degradation
introduced by the telescope diffraction pattern.
Simultaneously it keeps the atmospheric model as simple as possible. The atmospheric parameters returned by the code provide
the best fits to the Stokes profiles in the absence of the blurring
effect of telescope diffraction.

The inversion is performed on a denser spatial grid than that of the original data. The
resulting inverted parameters have a significantly improved spatial resolution
in most fits and appear to produce a more robust
inversion result than at the original pixel size \citep{vannoort2013}. Here we use a pixel size of $0.08''$/pixel,
a factor 2 smaller than in the original data, allowing structures
down to the diffraction limit of the telescope to be adequately
resolved. The returned parameters show 
small-scale structures
sharper than in the original data to the extent allowed by
noise (see \citet{vannoort2012,vannoort2013} for details and a discussion of the results of equivalent inversions of similar Hinode/SP data).

The inverted parameters are temperature $T$, magnetic field
strength $B$, field inclination relative to the line-of-sight $\gamma_{LOS}$, field azimuth $\phi$, line-of-sight velocity
$v_{LOS}$, and a micro-turbulent velocity $v_{MIC}$. All the free parameters
were allowed to vary at all three height nodes, which were placed
at $\log(\tau) = 0.0, -0.8$ and $-2.0$, respectively. The stratifications are then extrapolated linearly above $\log(\tau)=-2.0$ up to $\log(\tau)=-4.0$, and below $\log(\tau)=0$ down to $\log(\tau)\sim 1.3$.

The inversion returns very large field strengths, in excess of 7 kG, in about 200 pixels located near the umbra/penumbra boundary of  the center-side penumbra (see e.g. yellow markers on Figure \ref{fig:3b*}).
Figure \ref{fig:1d*} shows the observed Stokes profiles (dashed lines) in one of those pixels (yellow marker on the maps of Figure \ref{fig:1}). These profiles are highly complex since they exhibit large asymmetries and multi-lobed Stokes $V$ profiles, which causes their best-fits from the inversion to be not nearly as good as in most of the penumbral pixels. The SPINOR 2D best-fits to these profiles (solid red lines) feature very large line-of-sight velocities and magnetic field strengths 
at all three height nodes in order to reproduce the large wavelength separation in terms of the Zeeman splitting: $v_{LOS}\sim9.3$ km s$^{-1}$ and $B\sim8.3$ kG at $\log(\tau)=0$.

These unusually strong penumbral magnetic fields are not new, as they have been previously observed in supersonic penumbral downflows \citep{vannoort2013}. However,  they show up only when a spatially coupled inversion is performed and their reality needs to be confirmed with other techniques. To examine the reliability of the inversions in the pixels with very large field strengths is not the main aim of this work. We instead exclude in the present analysis all those pixels where the inversion gives $B>7$ kG.

Also, it is an intrinsic problem of inversions to specify errors in the fitted atmospheric parameters. Especially in the case of the 2D coupled inversions, the 
changes in the parameters of a single pixel severely affect the result, and therefore the error, of the neighboring pixels. This fact makes the computation 
of formal errors for a single pixel impossible. 
We stress that the main error of inversions is not the formal error in one pixel resulting from the minimization procedure, but is introduced by the choice of the correct model atmosphere.
The best error estimate therefore can be provided by  comparing the results from different model atmospheres. Such an analysis is beyond the scope of this work and will be presented in detail in another publication.
 Preliminary results of this analysis indicate that strong magnetic field values of 5 kG and even $\sim6$ kG represent a valid solution for the fit to the Stokes profiles in multiple models, and that models returning magnetic fields larger than 7 kG provide the best fit to the observed profiles.

Finally, after the inversion, the $180^{\circ}$ azimuthal ambiguity was resolved using the Non-Potential Magnetic Field Computation method \cite[NPFC;][]{Georgoulis2005}, which determines the non-potential component of the field to minimize the vertical current density $J_z$. 
The NPFC code also converts the values of magnetic field inclination and azimuth from the line-of-sight (LOS) frame into the local reference frame (LRF).
This step helps to determine the correct inclination and azimuth of the field in the LRF and is helpful for interpreting our results. However, we are aware that disambiguation techniques may not give reliable results at the small scales studied in this paper. Some of the reasons why such techniques may fail are: 1) the smallest structures we are studying have a horizontal dimension that is similar to the vertical corrugation of the $\tau=1$ surface. This invalidates the general assumption of disambiguation techniques that the field was measured on a flat surface. 2) Electric currents (e.g. current sheets at boundaries of filaments) are likely not properly resolved and thus may be underestimated. This may influence the results of techniques that aim to minimize the non-potential part of the field, such as the technique of \citet{Georgoulis2005}. 3) In the highly dynamic environment of the penumbra, with its waves, supersonic flows, chromospheric jets, etc., it is not clear if a minimization of the current really makes sense at the scales we are studying in the photosphere. These caveats must be borne in mind when considering the results in the LRF. 

\section{Results}\label{results}
In Figure \ref{fig:1}, we show  normalized maps of the Stokes parameters in the sunspot from a single SOT/SP scan. 
Except for the additional umbral-like feature in the bottom part of the maps, and its corresponding penumbral-like extension ($Solar$ $X\approx$ $[-666'',-634'']$,  $Solar$ $Y\approx$ $[-126'',-115'']$), which we do not study in this work,
the well-developed penumbra  surrounding the main umbra of this sunspot seems, at first sight, quite normal: it shows  quasi-radial filamentary structures all around the sunspot, and it is observed within a range of continuum intensities of $26-94\%$  that of the average quiet Sun.

However, a more careful inspection through the individual Stokes profiles at different places within the penumbra reveals an anomalous aspect: while it is possible to observe the normal EF (NEF), i. e., the photospheric absorption lines over the limb-side and the disk-center-side penumbrae are redshifted and blueshifted, respectively, indicating motions away from and towards the observer (see, e. g., Stokes profiles in Figures \ref{fig:1a*} and \ref{fig:1b*}, respectively), strong redshifts are also observed over a broad portion of the center-side penumbra (see, e. g., Stokes profiles in Figure \ref{fig:1c*}), indicating a counter EF (CEF).

The existence of the CEF in the center-side penumbra becomes clearer after the inversions, since most of the observed line profiles from the center-side penumbra show the same magnetic polarity (i. e. same signs of the Stokes $V$ profiles) and  differ from those in the NEF region only in more subtle ways such as in the line shifts and asymmetries. Nonetheless, the Stokes maps in Figures \ref{fig:1a},  \ref{fig:1b} and  \ref{fig:1c} suggest that the penumbral filaments in the CEF region have a slightly different orientation (not completely radially oriented) compared to that in the undisturbed other parts of the penumbra (which contain more radially oriented filaments). 

\begin{figure*}[htp!]
    \centering
    \includegraphics[width=\textwidth]{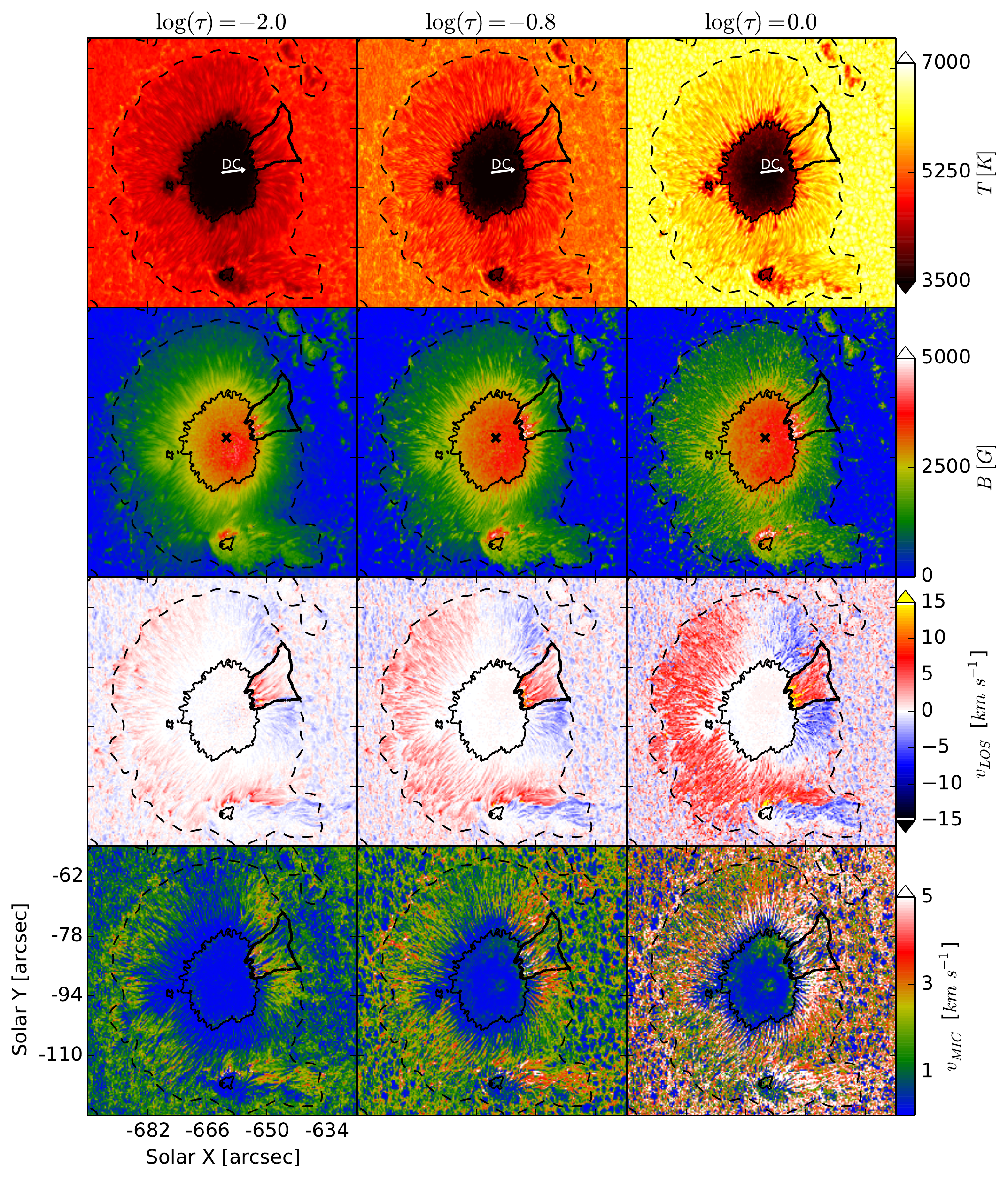}
    \caption{SPINOR 2D inverted parameters  at three photospheric layers. 
\textit{From left to right:} $\log 
(\tau)=-2.0$, $-0.8$ and $0$. \textit{From top to bottom:}  temperature $T$ (K); magnetic field intensity $B$ (G); line-of-sight velocity $v_{LOS}$ (km s$^{-1}$); and the micro-turbulence velocity $v_{MIC}$ (km s$^{-1}$).
In all maps, the black thick contour encloses a penumbral region where a counter EF is observed.
The white arrows on the  temperature maps point towards the disk center.
Black crosses on the  $B$ maps indicate the location of a local maximum of the umbral field strength at $\log  (\tau)= 0$.
Also, the color bar scale is sometimes saturated. }\label{fig:2}
\end{figure*}

Figure \ref{fig:2} shows some of the resulting maps of the  physical parameters inverted with SPINOR 2D at three continuum optical depth levels, $\log(\tau)=-2.0, -0.8$ and $0$ displayed from left to right. From the $v_{LOS}$ maps (third row), it is possible to observe the NEF, i. e., the limb-side and center-side penumbrae are redshifted and blueshifted, respectively. 
 Additionally, as indicated  by the black thick contours,  the CEF is observed over a rather large area of the center-side penumbra, which mostly contains positive $v_{LOS}$ values indicating motions away from the observer. Such motions, because of their center-side location close to the symmetry line (line connecting the sunspot's center with the disk-center), represent either downward motions of material or inflows directed from the outer penumbra towards the umbra of the sunspot. 

\begin{figure*}
    \centering
    \includegraphics[width=\textwidth]{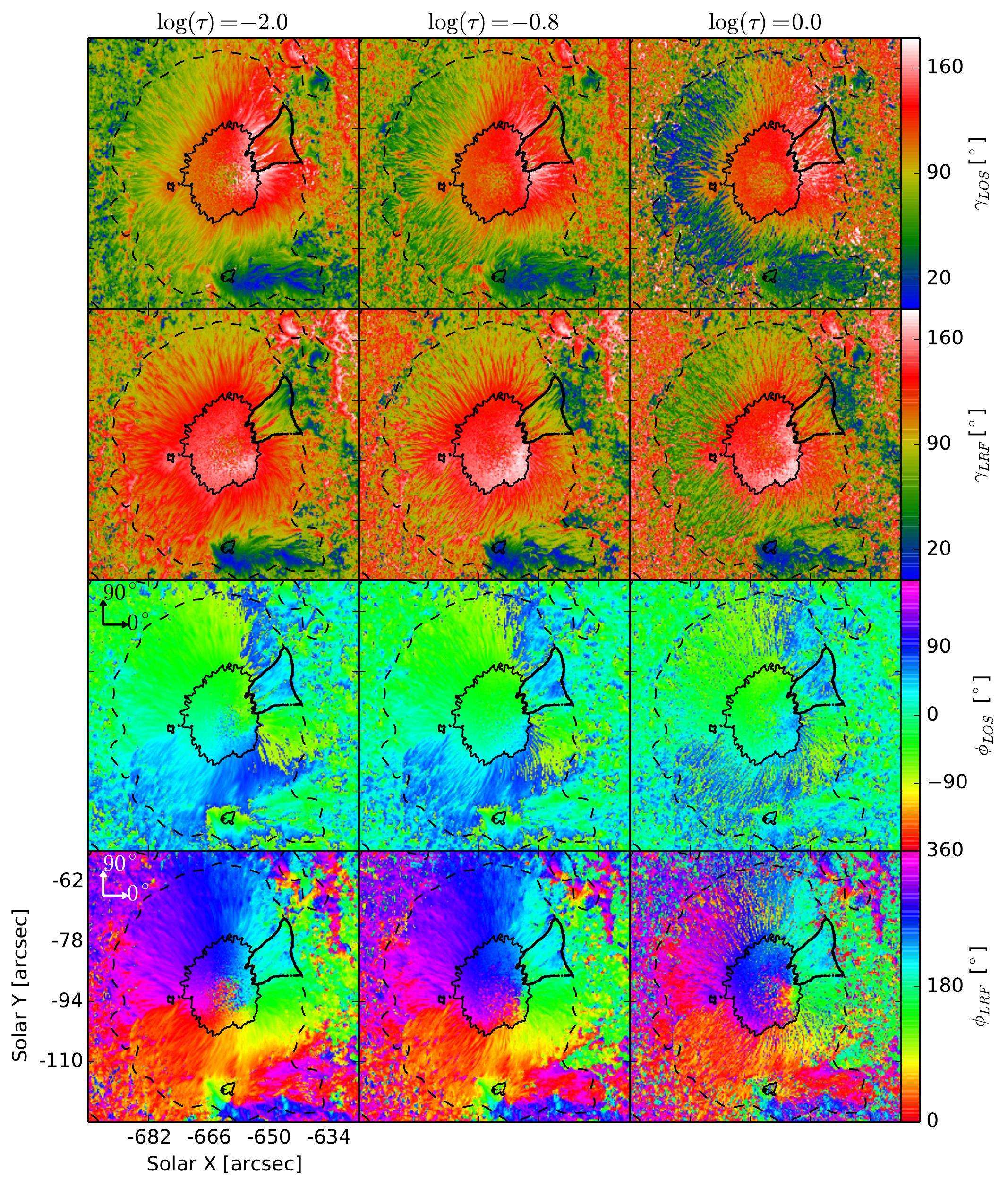}
    \caption{Field inclination and field azimuth from SPINOR 2D inversions  at three photospheric layers. 
\textit{From left to right:} $\log 
(\tau)=-2.0$, $-0.8$ and $0$. \textit{From top to bottom:}  the  field inclination angle in the line-of-sight direction $\gamma_{LOS}$ ($^{\circ}$); field inclination in the local-reference-frame $\gamma_{LRF}$ ($^{\circ}$)  after the disambiguation of the field azimuthal angle;
 the ambiguous field azimuthal angle  in the line-of-sight direction $\phi_{LOS}$ ($^{\circ}$); and the disambiguated field azimuthal angle  in the local-reference-frame  $\phi_{LRF}$ ($^{\circ}$). The contours are the same as in Figure \ref{fig:2}.
}\label{fig:2*}
\end{figure*}

\begin{figure*}[htp!]
 \centering
 \begin{subfigure}[b]{0.5\textwidth}
\centering
        \includegraphics[width=\textwidth]{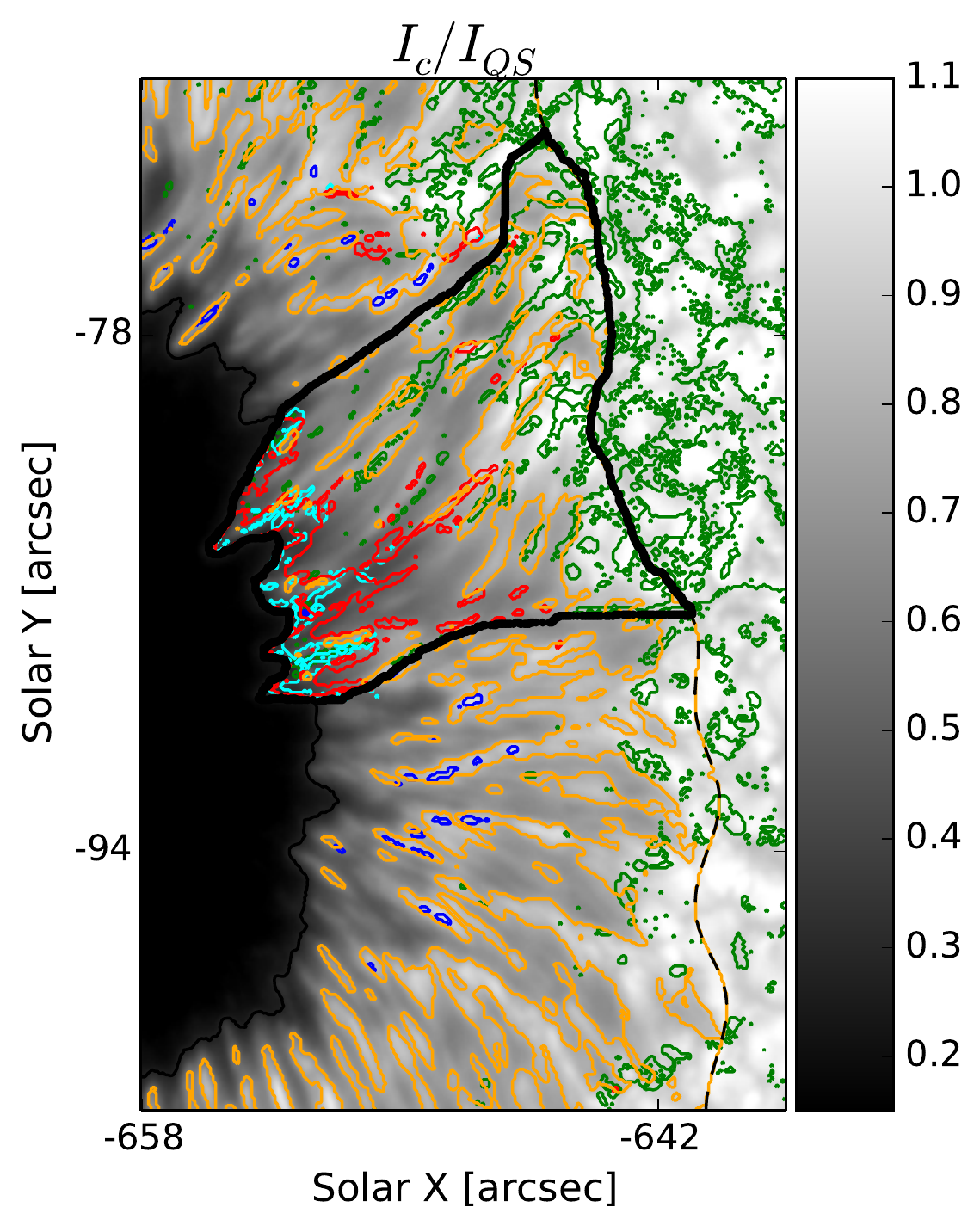}
        \caption{}
	\label{fig:3a*}
     \end{subfigure}%
    ~
    \begin{subfigure}[b]{0.5\textwidth}
        \centering
        \includegraphics[width=\textwidth]{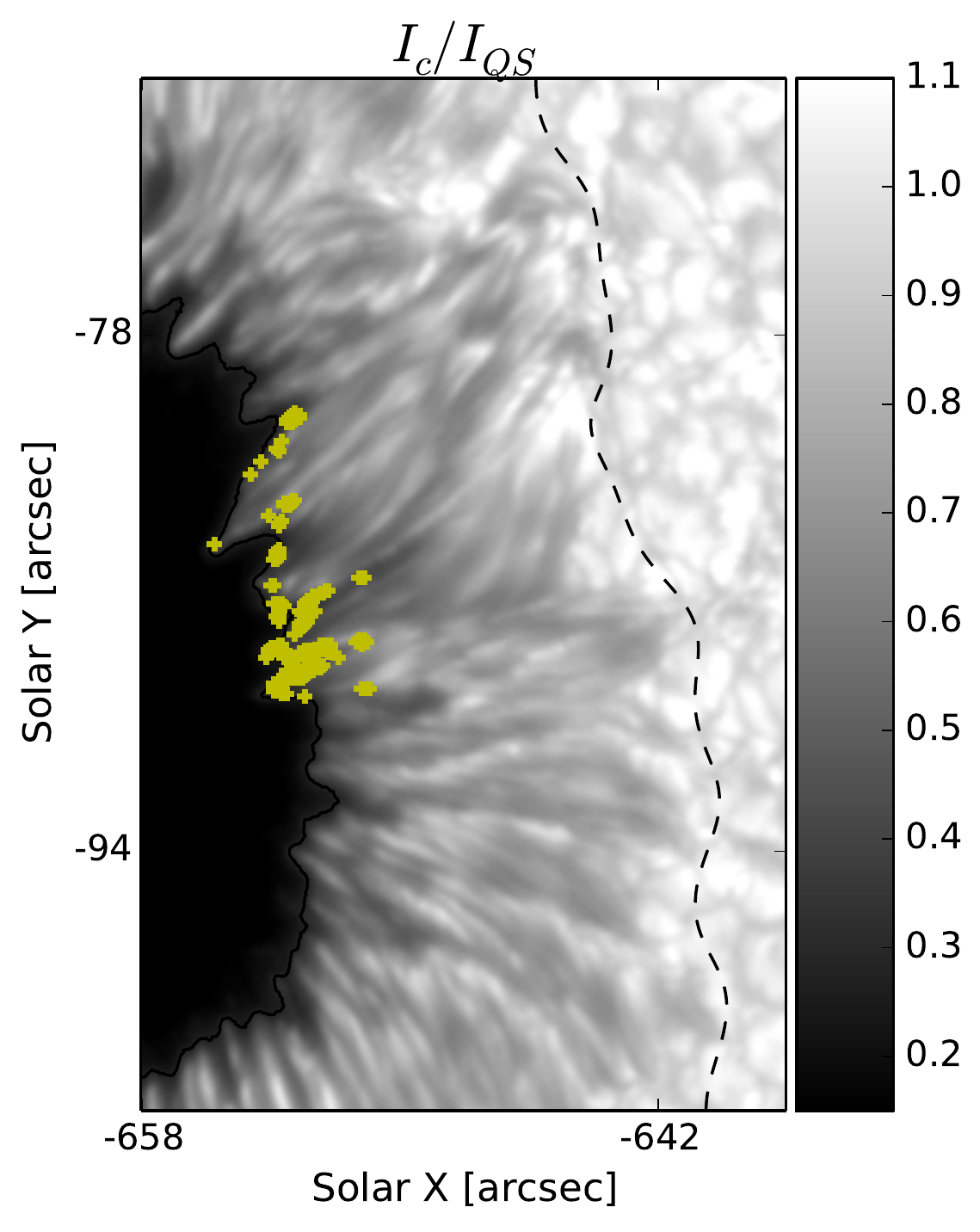}
        \caption{}
	\label{fig:3b*}
    \end{subfigure}%

     \caption{Continuum intensity maps of the center-side penumbra. (a) The black thick contour encloses the CEF region. The colored contour lines enclose the regions with line-of-sight velocities exceeding $9$ km s$^{-1}$ (red), regions with $v_{LOS}<-9$ km s$^{-1}$ (blue), regions where $B>5$ kG (cyan) and regions where $T>6000$ K (orange) within the penumbra.  The green contours enclose regions where  $\gamma_{LRF}<90^{\circ}$ and the amplitude of Stokes $V$ is at least $10 \sigma$.
All contouring was made at $\log(\tau)=0$.
 (b) Yellow marks indicate the location of pixels where the SPINOR 2D inversions return $B\geq7$ kG at $\log(\tau)=0$.}
\label{fig:3*}
\end{figure*}

Both, the NEF and the CEF are observed at all three selected atmospheric layers.
The largest blueshifts are observed at the deepest observable layer ($\log(\tau)=0$) in the center-side NEF penumbra, with associated velocities (in the line-of-sight direction) that reach values of up to $v_{LOS} \sim-15$ km s$^{-1}$. 
The largest redshifts are located at the innermost part of the CEF penumbra (towards the umbra-penumbra boundary). These redshifts also become stronger with depth, corresponding to extreme velocities ($v_{LOS}>20$ km s$^{-1}$) and  exceeding even the ones found by \citet{vannoort2013} in peripheral downflows within sunspot penumbrae ($v_{LOS}\sim22$ km s$^{-1}$).

The inverted maps in Figure \ref{fig:2} do not show a clear difference in temperature between the part of the penumbra showing a NEF and the one harboring the CEF, in any of the three atmospheric layers. However,  at all three layers, the magnetic field strength
appears to be larger at the inner part  of the CEF penumbra than in any part of the NEF penumbra. Moreover, the inversions give values $B>5$ kG in some regions located in the innermost part of the CEF penumbra  (see cyan contours in Figure \ref{fig:3a*}). In particular, those regions contain pixels, located exactly at the umbral/penumbral boundary (see yellow markers in Figure \ref{fig:3b*}), where $B>7$ kG according with the inversions. 
Such field strengths, apart from being extremely high compared with typical penumbral field strengths, are also stronger than the umbral field itself (the umbral field  at $\log(\tau)=0$ has a local maximum close to the umbra's geometric
 center as indicated by black crosses in the magnetic field maps of Figure \ref{fig:2}, where $B\sim3.9$ kG, although some  pixels close to the center-side umbral boundary reach values up to $B\sim4.2$ kG). 

Figure \ref{fig:2*} shows the field inclination $\gamma$ and azimuth $\phi$ in both, the LOS frame and  LRF.
The transformation of the magnetic field vector to the LRF suggests that the CEF region has different magnetic properties than the rest of the penumbra given that, at all three heights, the magnetic field lies more horizontally within the enclosed penumbral sector harboring the CEF, even changing polarity in the outer penumbra at $\log(\tau)=-0.8$ and $-2$.
The maps of $\gamma_{LRF}$  also show a number of patches of opposite polarity  to that of the sunspot's umbra, in the center-side penumbra at all three optical depth levels. They  are found mainly at the outer penumbral boundary in both, the  center-side NEF and the CEF regions, although there are  many more opposite polarity  patches concentrated closer to the outer penumbra in the CEF region. 
We can also see this in Figure \ref{fig:3a*}, where green contours have been placed on the continuum intensity map for the center-side penumbra enclosing regions where $\gamma_{LRF}<90^{\circ}$  at $\log(\tau)=0$ and the amplitude of Stokes $V$ is at least $10 \sigma$.

The overall variation of $v_{LOS}$, $T$ and $B$ with height in the CEF part of the penumbra is similar to that of the NEF part: they all increase with depth (see Figure \ref{fig:2}). Nevertheless, in the anomalous region, $v_{LOS}$ and $B$ are observed to reach more extreme values. In particular, in the inner penumbra of the CEF region  $v_{LOS}>20$ km s$^{-1}$ and $B>5$ kG are observed at $\log(\tau)=0$.
Such high values of $v_{LOS}$ and $B$ are atypical in the penumbrae of sunspots, with the closest values to these being those reported by \citet{vannoort2013}. Even more, field strengths  $B >7$ kG are unusual even for a dark umbra \cite[see][]{Livingston2006}. We do not include the pixels where $B >7$ kG in the following analysis. Instead, we plan to conduct a more detailed study on the reality of unusually strong fields found in sunspots penumbrae. This is beyond the scope of this paper and will be the topic of a future work.

\begin{figure}
   \centering
 \resizebox{\hsize}{!}{  \includegraphics[width=\hsize]{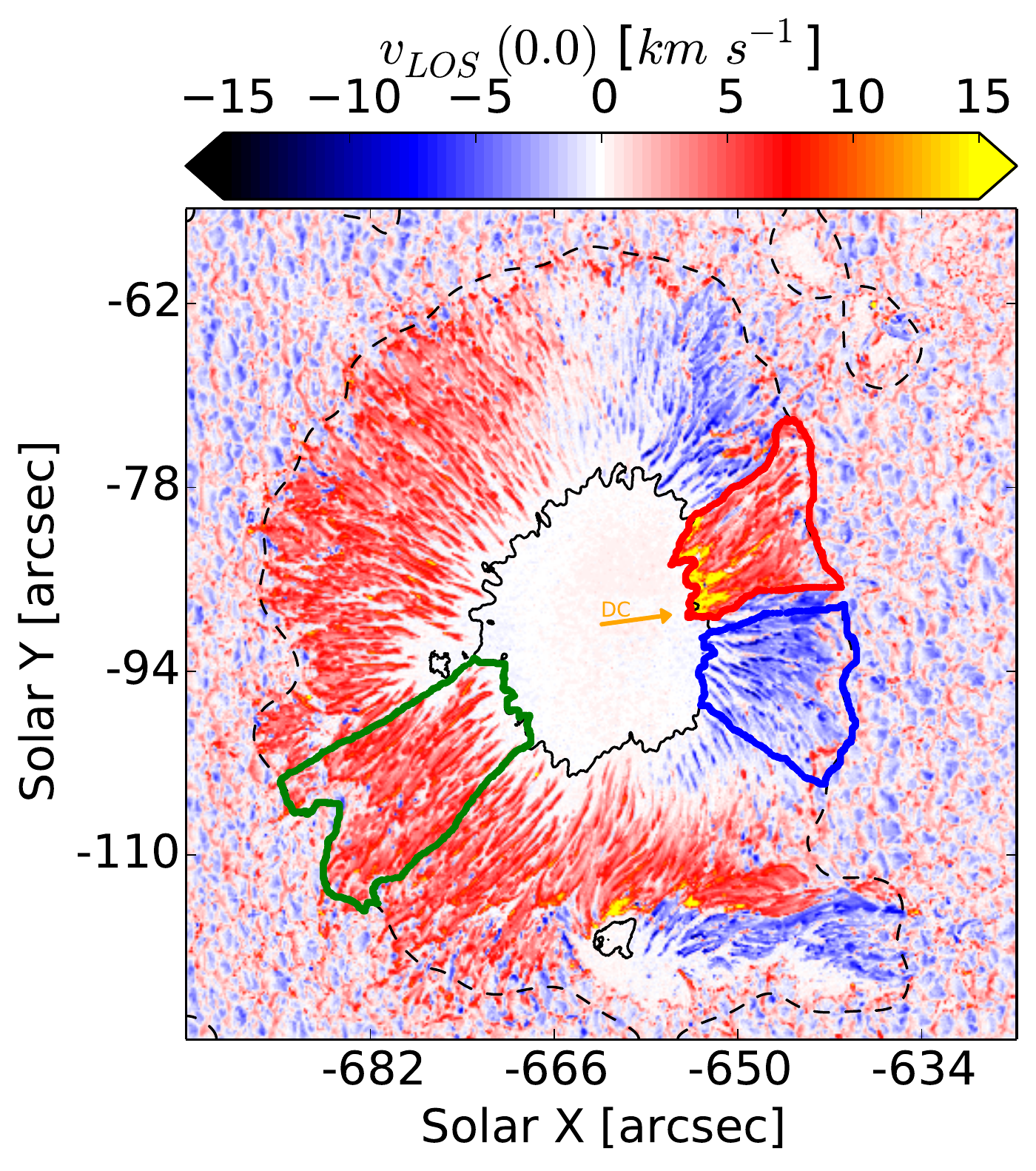}}
      \caption{ Three regions of  interest are highlighted on the $v_{LOS}$ map at $\log(\tau)=0$:  region with CEF (red), center-side region with NEF (blue) and limb-side region with NEF (green).
 }
         \label{fig:4}
   \end{figure}

\subsection{Filament Selection}

In order to compare the differences and/or similarities between the fine-scale structures related to the CEF and the NEF, we  investigate now the generic properties of the filaments that populate the three  penumbral regions identified in Figure \ref{fig:4}. For this, we manually selected 6 filaments from each of these penumbral regions.

The individual filaments were selected based on selection criteria  introduced by \citet{Tiwari2013}, but adapted to a sunspot located off the disk center ($\theta \approx 47^{\circ}$). In our selection criteria, we use the temperature, the field inclination angle and the line-of-sight velocity, at $\log(\tau)=0$, as follows: (1) the heads of the filaments are identified by relatively warm upflows and nearly vertical magnetic fields; (2) the bodies of the filaments are characterized by more horizontal fields and by the signature of the Evershed flow in $v_{LOS}$ due to the large heliocentric angle of the sunspot; (3) the tails of the filaments are localized in regions of concentrated downflows and of nearly vertical fields of opposite polarity to the umbra.

The selection procedure was applied manually by placing points
along the central axis of the filament. A line
connecting these points was then computed using a bi-cubic
spline interpolation. The path defined in this way defines the possibly curved
axis, or spine of the filament and was used in the de-stretching and  length
normalization of the filament.

\begin{figure}
   \centering
\resizebox{\hsize}{!}{\includegraphics[width=\hsize]{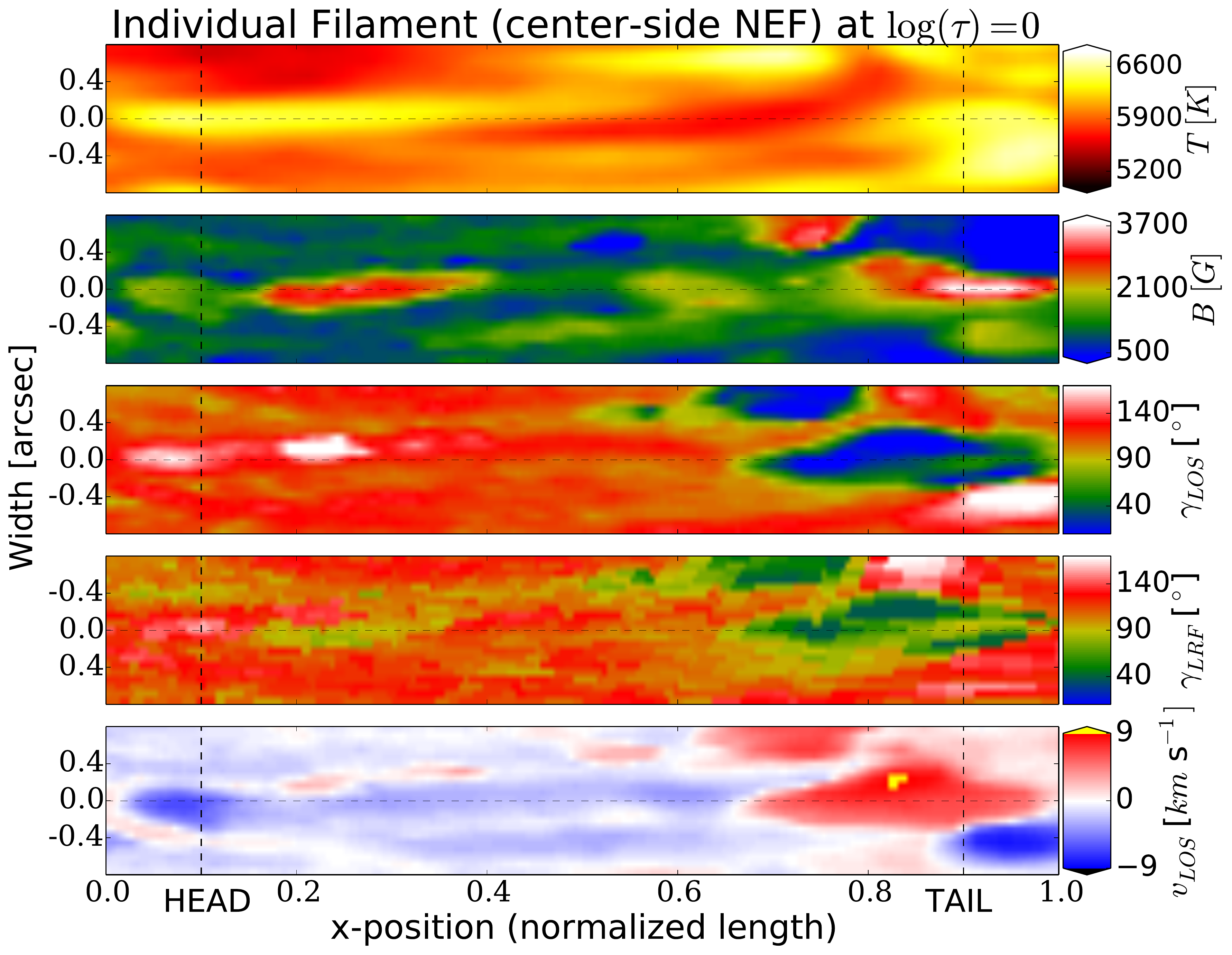}}

\resizebox{\hsize}{!}{\includegraphics[width=\hsize]{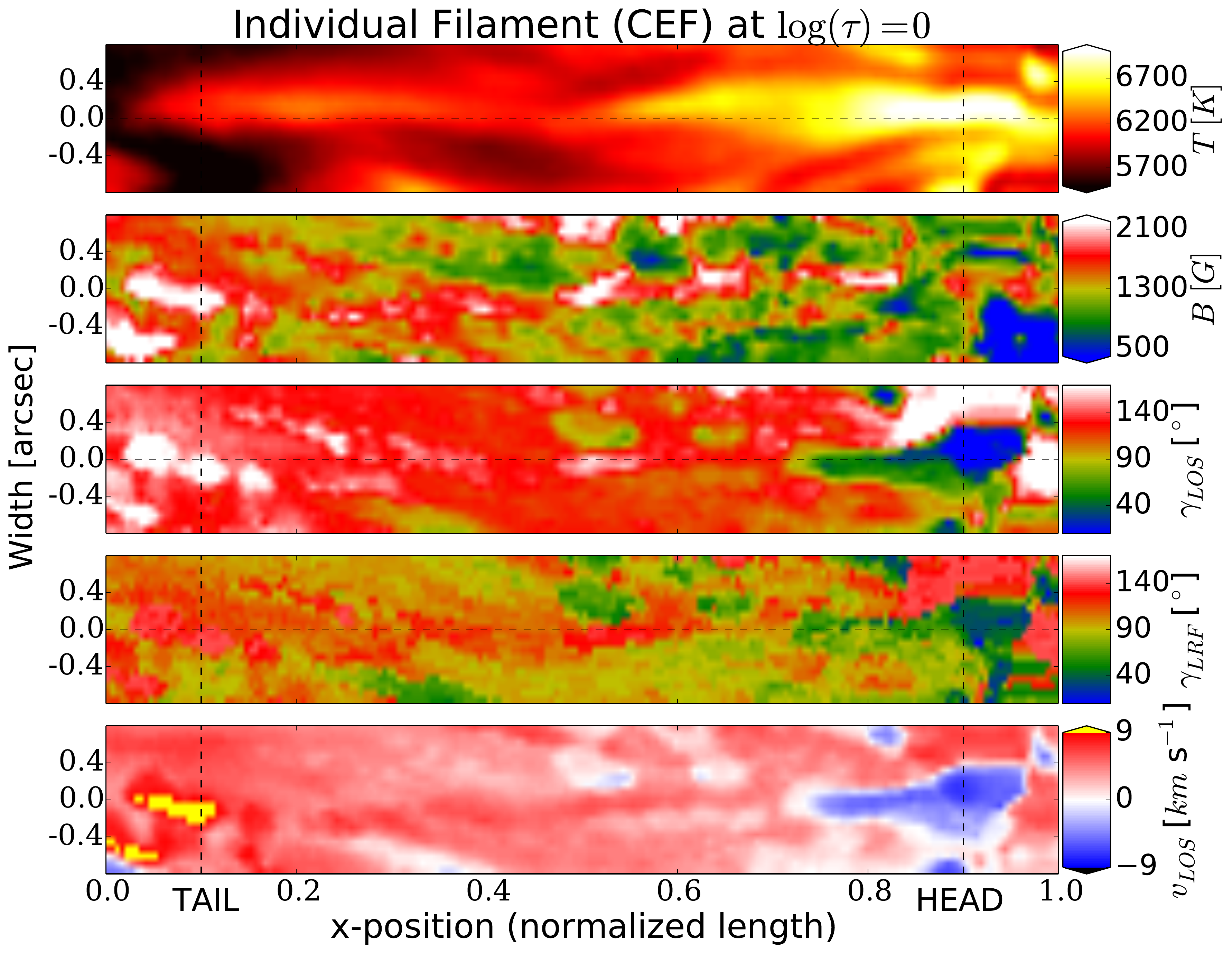}}
      \caption{Individual filaments carrying a NEF (top) and a CEF (bottom) 
in their de-stretched and scaled form as seen at $\log(\tau)=0$. Subplots show, from top to bottom: the temperature $T$,  magnetic
field strength $B$, field inclination in the line-of-sight $\gamma_{LOS}$ and in the local refrence frame $\gamma_{LRF}$, and the  line-of-sight velocity $v_{LOS}$.  The vertical dashed lines indicate transversal cuts close to the filament's endpoints: inner or closer to the umbra (left) and outer or closer to quiet sun (right). Head refers to the footpoint harboring an upflow, tail to the footpoint showing a downflow. The horizontal dashed lines indicate the de-stretched and length normalized axis of the filament. Note that color bars for a given parameter do not always have the same
range for both maps of a given physical parameter and some of them have been saturated to highlight some patterns. 
              }
         \label{fig:9}
   \end{figure}

To perform the de-stretching and length normalization,
200 equidistant points were placed along the axis of each filament,
after which ten points on a line perpendicular to the tangent
are placed at intervals of one pixel ($0.08''$) on each side
of the path. A cubic interpolation of the inverted parameters in
these points results in a de-stretched, de-rotated and length normalized
filament, as shown in Figure \ref{fig:9} for two individual filaments, one from the center-side NEF and one from the CEF penumbral regions.

With the method described above, a total of
18 penumbral filaments were selected and de-stretched, 6 inside each of the three  penumbral regions highlighted in Figure \ref{fig:4}.  
 Our original aim was to select equal numbers of filaments located in the inner, middle, and outer parts of the penumbra within each highlighted region. 
Due to geometrical constraints introduced when using  $\gamma$  as a selection parameter, it turned out to be most reliable to identify  filaments in the inner penumbra in the limb-side NEF region. Likewise, all selected filaments in the center-side NEF region lie in the outer penumbra. In contrast, the filaments in the CEF region turned out to be much longer; most of them extend over the whole penumbral width, i. e., they originate close to the boundary between penumbra and quiet Sun (i. e. this is where the upflow in these filaments is located) and end at the edge of the umbra.
Such a positional difference between the filaments from each group should, however, not prevent us from performing a qualitative comparison between the filaments from the 3 regions, since, according to \citet{Tiwari2013}, all filaments have essentially the same structure independently of their location within the penumbra, with the biggest differences happening in their surroundings.

\begin{table*}

\begin{center}
    \begin{tabular}{ c|c c c c c }
    \hline
    & $l_{mean}$ (arcsec)& $l_{std}$
 (arcsec)& $l_{median}$ (arcsec)& $w$  (arcsec)&$l/w$\\ \hline
    NEF (center-side)& 5.2&1.4&5.0&0.56&9.29\\ 
    NEF (limb-side)& 4.9&1.3&4.5&0.80&6.13\\ 
    CEF  & 9.4&2.7&11.0&0.96 &9.79\\
    \hline 
	\end{tabular}
\end{center}

 \caption{Geometrical properties of the filaments: mean length ($l_{mean}$), standard deviation ($l_{std}$) and median values ($l_{median}$) for each filament set,  average filament width ($w$)  and the corresponding length-to-width ratio ($l/w$).} \label{tab:2}
\end{table*}

\begin{figure}
   \centering
\resizebox{\hsize}{!}{   \includegraphics[width=0.8\hsize]{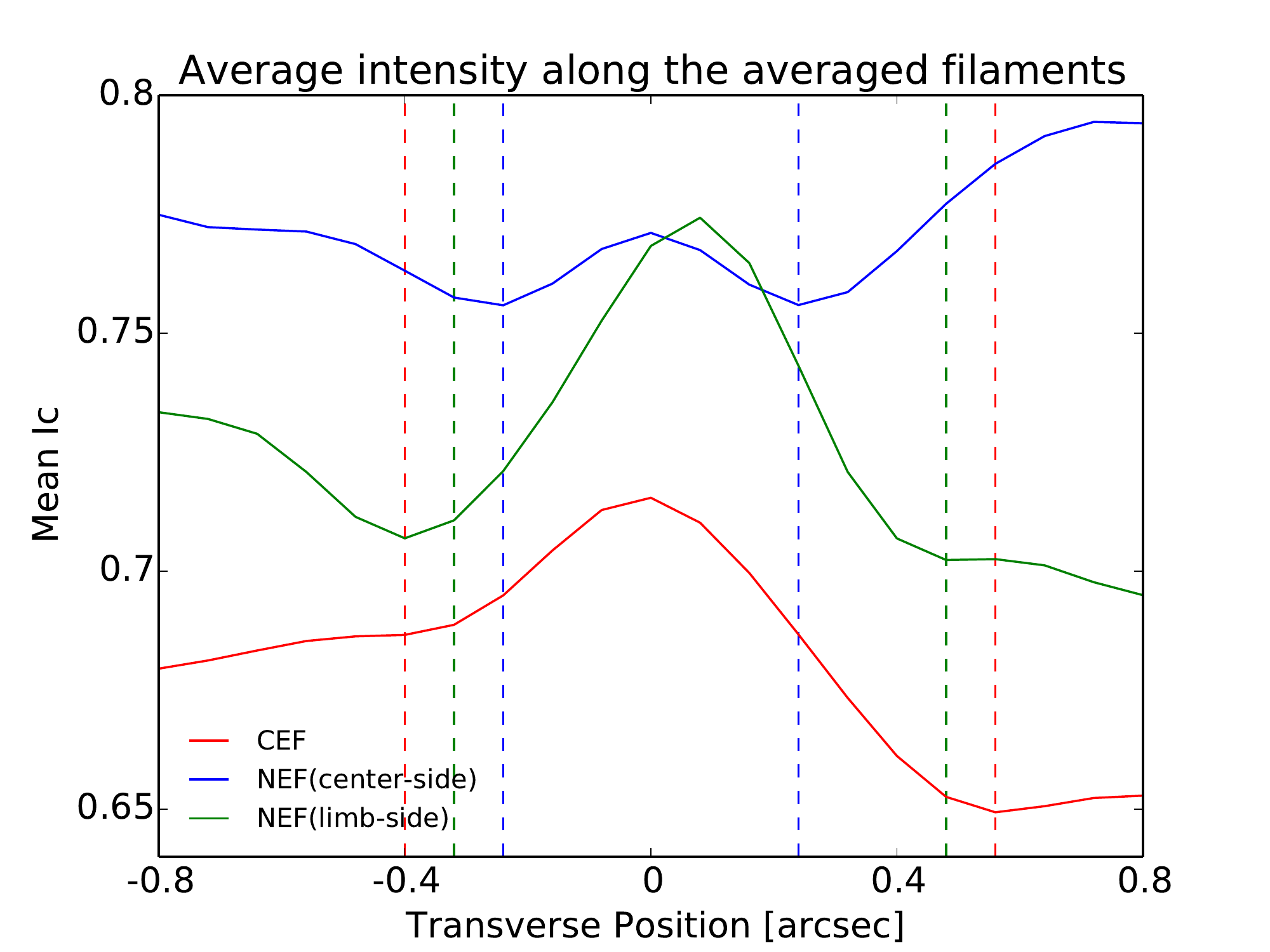}}
      \caption{Transversal continuum intensity profiles of the average filaments. The plotted $I_c$ is averaged along the axes of the 3 averaged filaments: CEF (red line) , center-side NEF (blue line) and limb-side NEF (green) regions.
              }
         \label{fig:11}
   \end{figure}

Table \ref{tab:2} shows the mean, the standard deviation and the median values of the length distributions for each of the selected filament sets in the penumbra. 
Strikingly, the CEF-carrying filaments are nearly twice as long as those hosting the NEF.

Given the range of lengths of filaments from each selected region, they are rescaled prior to averaging. It is important to keep in mind that while averaging, the variation in the length
will cause some properties to be re-scaled along the filament.
Nonetheless, the generic physical properties and structure,
reflected by the average of such filaments, should be valid for
filaments of all sizes.

Thus, after de-stretching and length normalizing all the selected filaments  we separately averaged each 6-filament set in order to highlight their common characteristics, producing a "standard" filament for each of the three selected parts of the penumbra.
This reduces the large variability in the appearance of the filaments caused by interactions with the environment as well as due to their intrinsic variability.
 Thus we finally just analyzed the above three standard or averaged filaments, one for each penumbral region marked in Figure \ref{fig:4}.

The width of the filaments is, unlike their lengths, not re-scaled. 
The  width values shown in Table \ref{tab:2}  were computed by averaging the
continuum intensity  along the axes of the 3 averaged  filaments. As shown in Figure \ref{fig:11}, they
come out to be roughly $0.6''$, $0.8''$ and $1''$ for the center-side NEF, limb-side NEF and CEF regions, respectively. Thus, from the average of the measured lengths of
the individual filaments, we arrive at a length-to-width ratio of
approximately $9$,  $6$ and $10$ for the center-side NEF, limb-side NEF and CEF regions, respectively.

\subsection{Qualitative picture of filaments}

Figure \ref{fig:12} depicts a selection of properties of each of the three
averaged filaments, representing the
selected penumbral region (center-side NEF, limb-side NEF and CEF), at $\log(\tau)=0$. We concentrate on $\log(\tau)=0$ since the important differences and similarities between the averaged filaments are best seen there.

By comparing plots in Figure \ref{fig:12} with similar images of the individual filaments (e. g. Figure \ref{fig:9}) we can see that, averaging 6 filaments in each of the three groups is sufficient to clearly highlight the common features of the filaments and to suppress the largest individual fluctuations.

\subsubsection{center-side and limb-side NEF filaments}
\begin{figure}[htp!]

   \centering
 \resizebox{\hsize}{!}{  \includegraphics{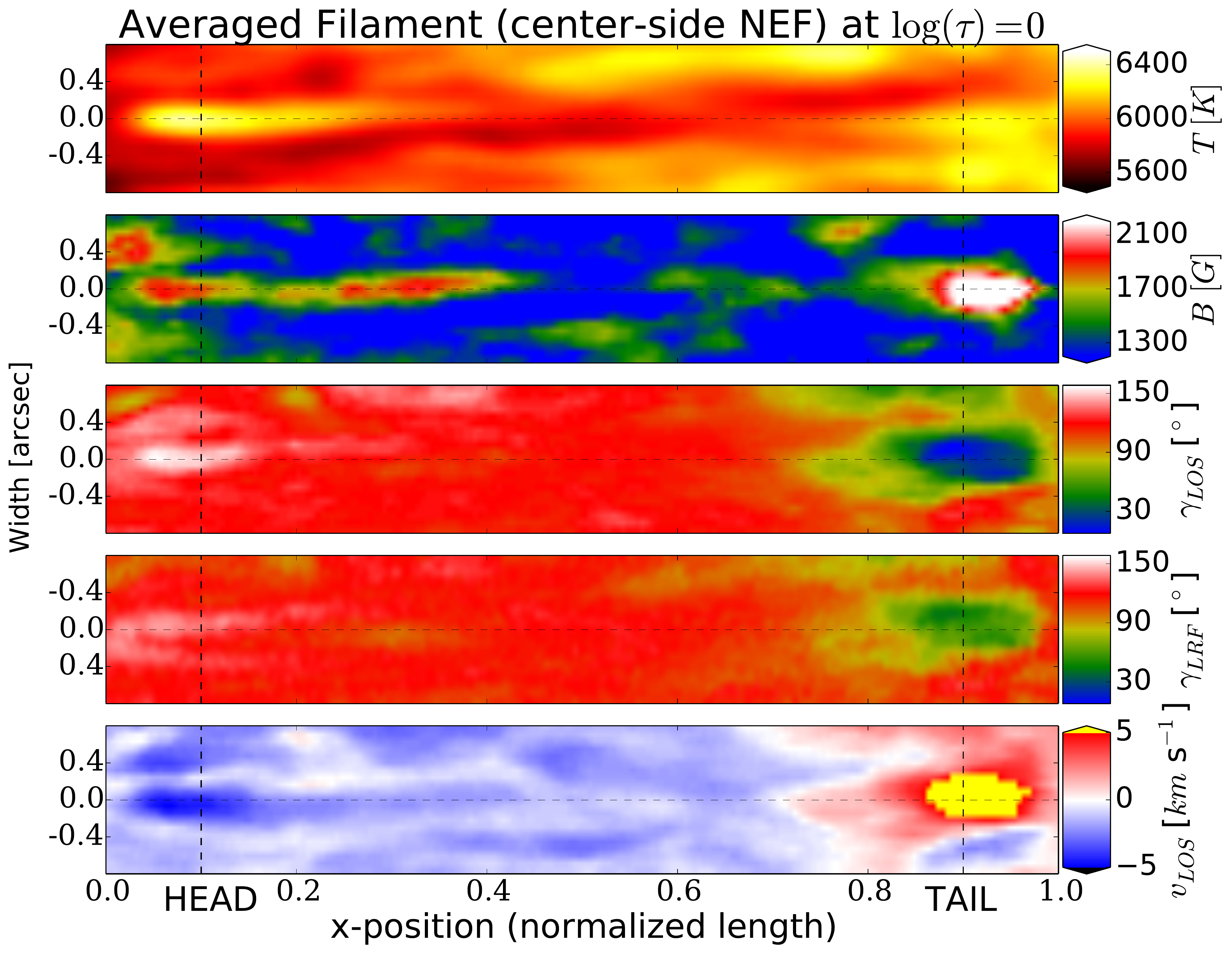}}
\resizebox{\hsize}{!}{\includegraphics{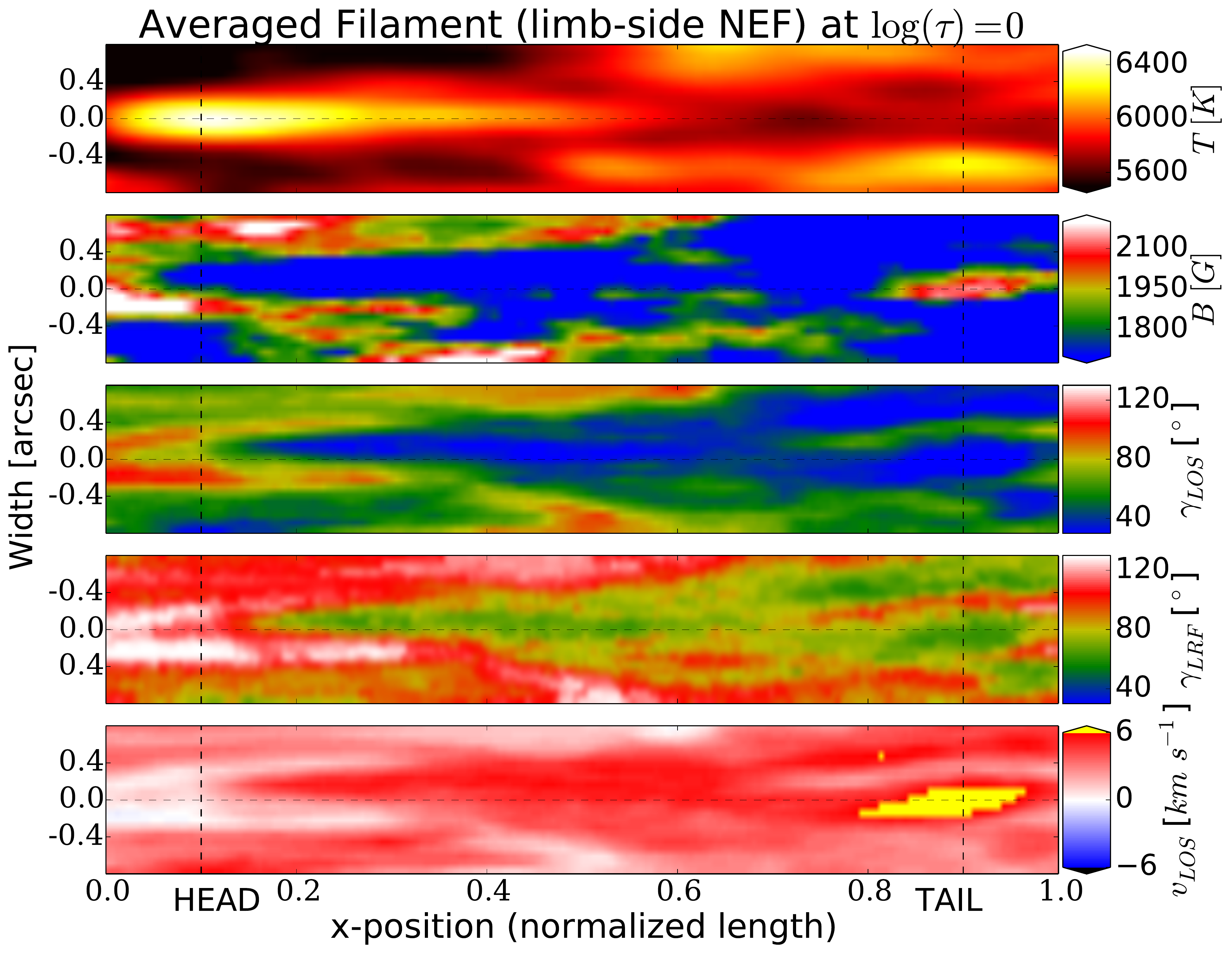}}
\resizebox{\hsize}{!}{\includegraphics{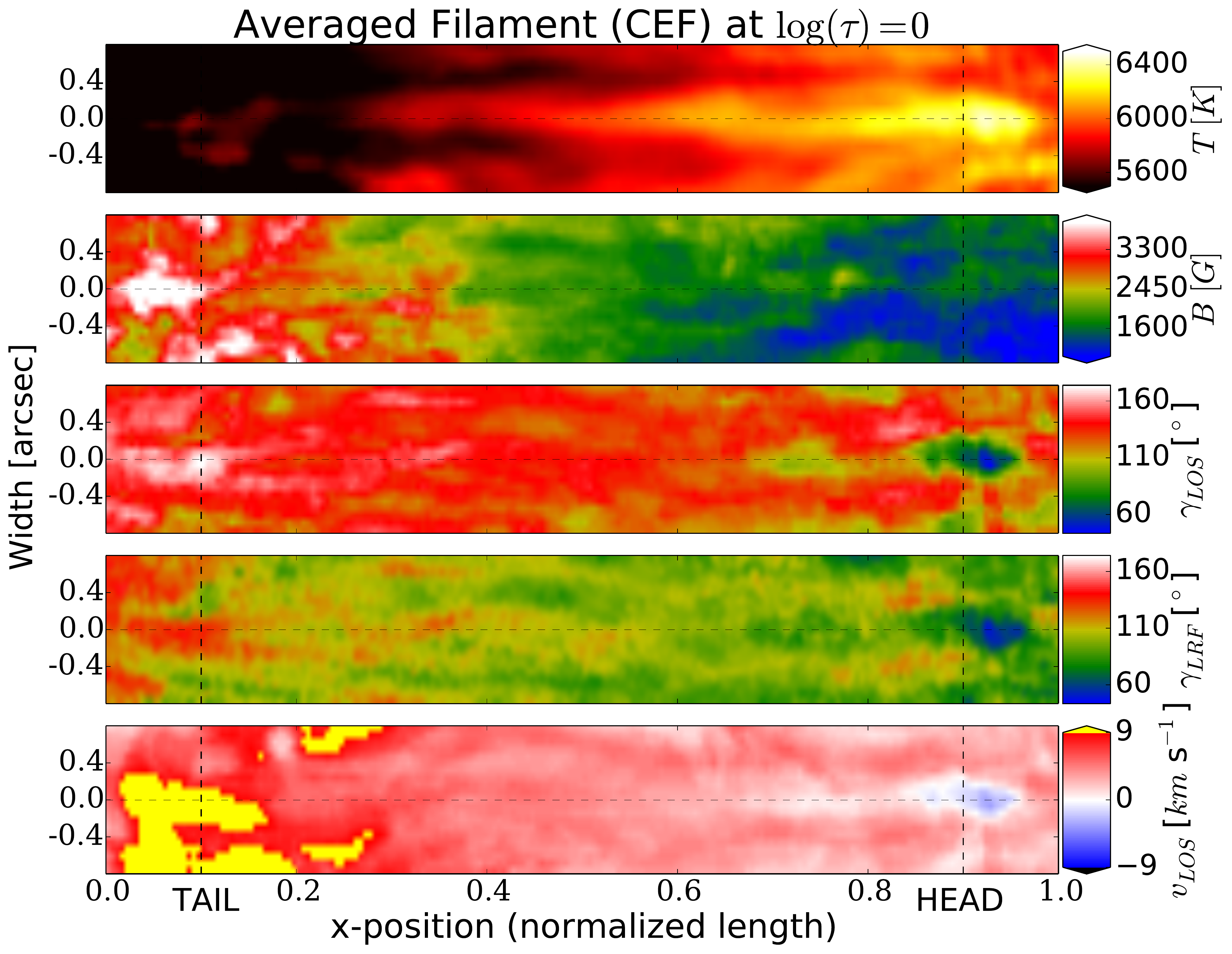}}
      \caption{Average filaments from the center-side  NEF (upper set of panels), limb-side NEF (middle panels) and  CEF (lower panels) region.  Subplots, i.e. the sets of 5 panels depicting filaments of a particular penumbral region, are in the same format as plots in Figure \ref{fig:9}. Note that the head of CEF is on the right, while the NEFs are plotted with their heads on the left, i. e. the filaments are plotted such that the part closer to the umbra is on the left. 
        }
         \label{fig:12}
   \end{figure}

Figure \ref{fig:12} shows that the average filament from the center-side NEF region starts with a strong peak in temperature ($\sim 6400$ K in the head). The temperature gradually decreases along the body of the filament and increases again towards the tail, up to $\sim6300$ K.
The magnetic field strength also decreases from the head towards the body from almost 1.9 kG to $\sim1.3$ kG, but then shows a local large strengthening at the very tail up to almost $2.5$ kG.

The LOS inclination angle suggests that the head of the center-side filament has a downward pointing field,  a more horizontal but slightly downward pointing field along the body (i. e. with the same polarity as the umbra), while the tail displays an opposite field polarity to that observed at the head, i. e. upward pointing field lines. This field configuration is maintained after transforming $\gamma$ to the LRF, i. e., the transformation gives $\gamma_{LRF} \sim 140^{\circ}$ in the head, $\sim 110^{\circ}$ along the body and $\sim60^{\circ}$ in the tail.

\begin{figure*}[htp]
   \resizebox{\hsize}{!}
            {\includegraphics[width=0.5\textwidth]{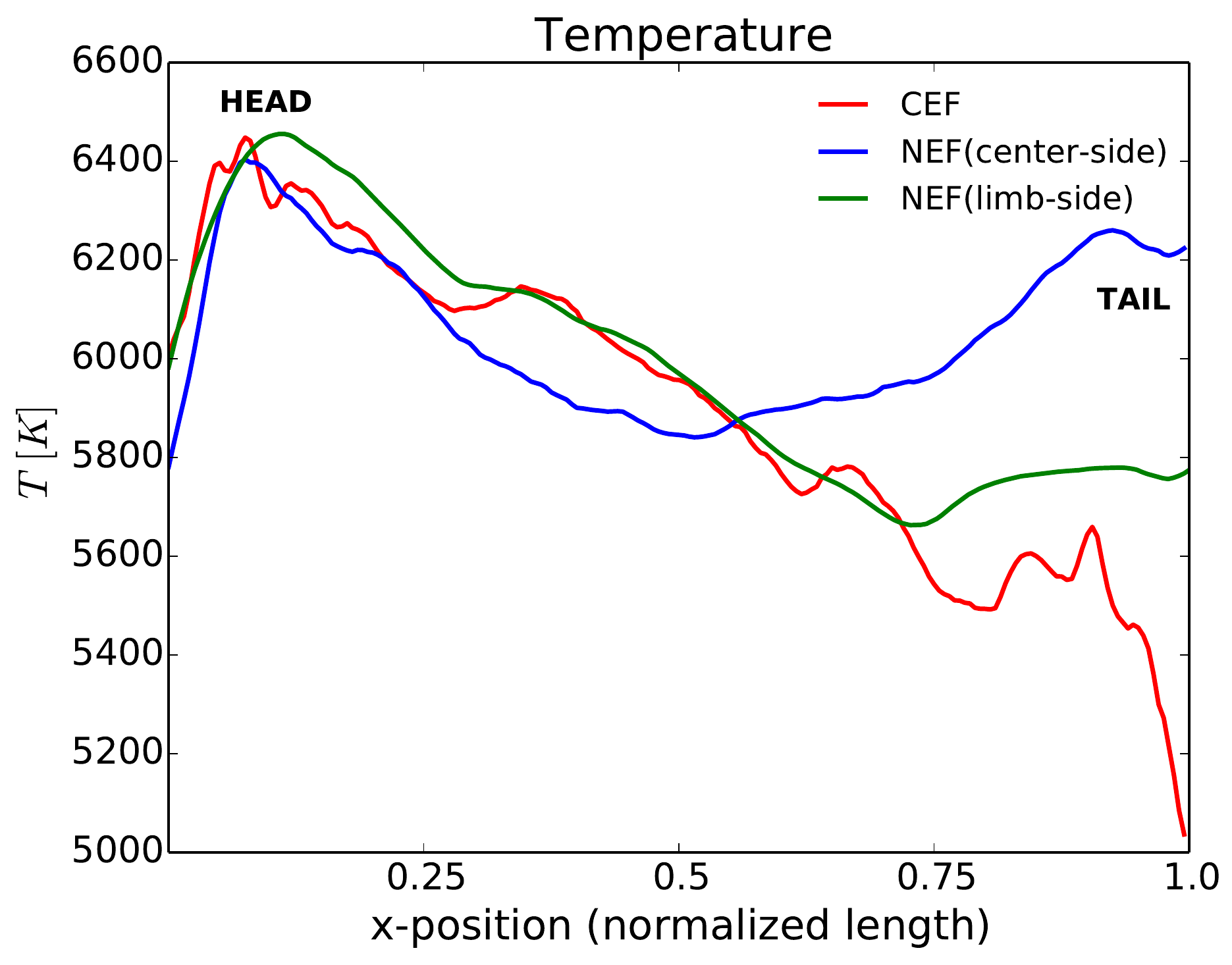}
 \includegraphics[width=0.5\textwidth]{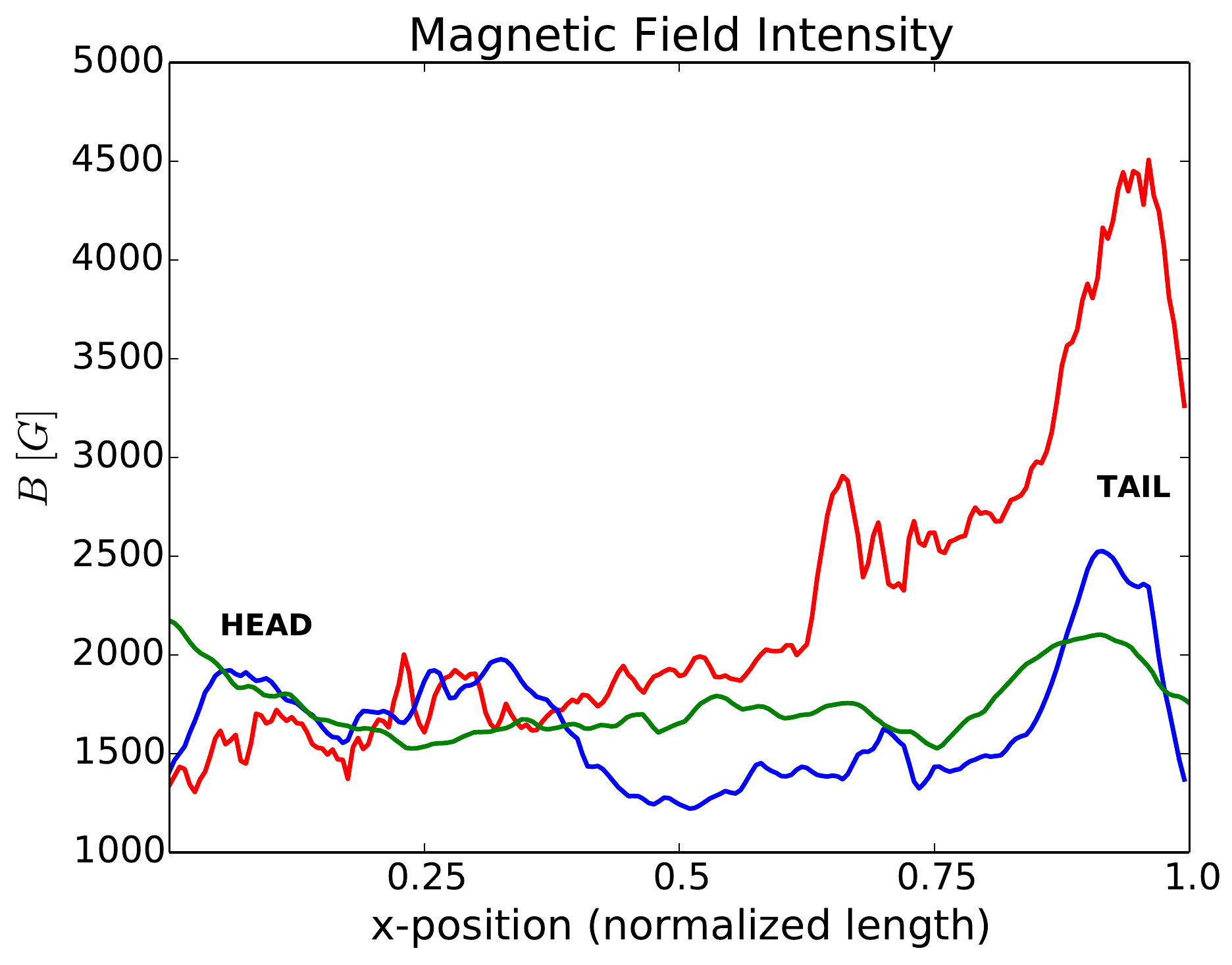}

}

\resizebox{\hsize}{!}
            {\includegraphics[width=0.5\textwidth]{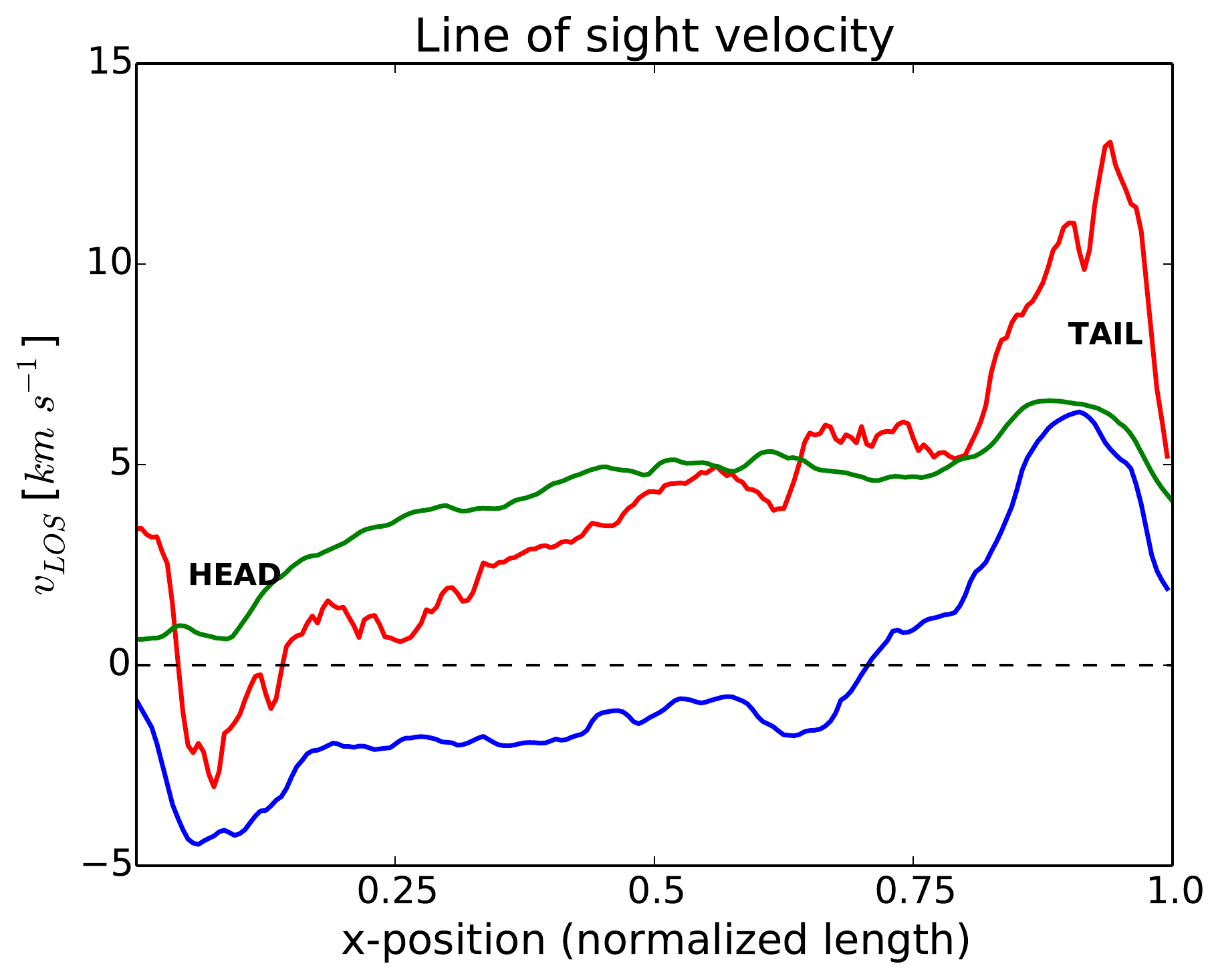}
\includegraphics[width=0.5\textwidth]{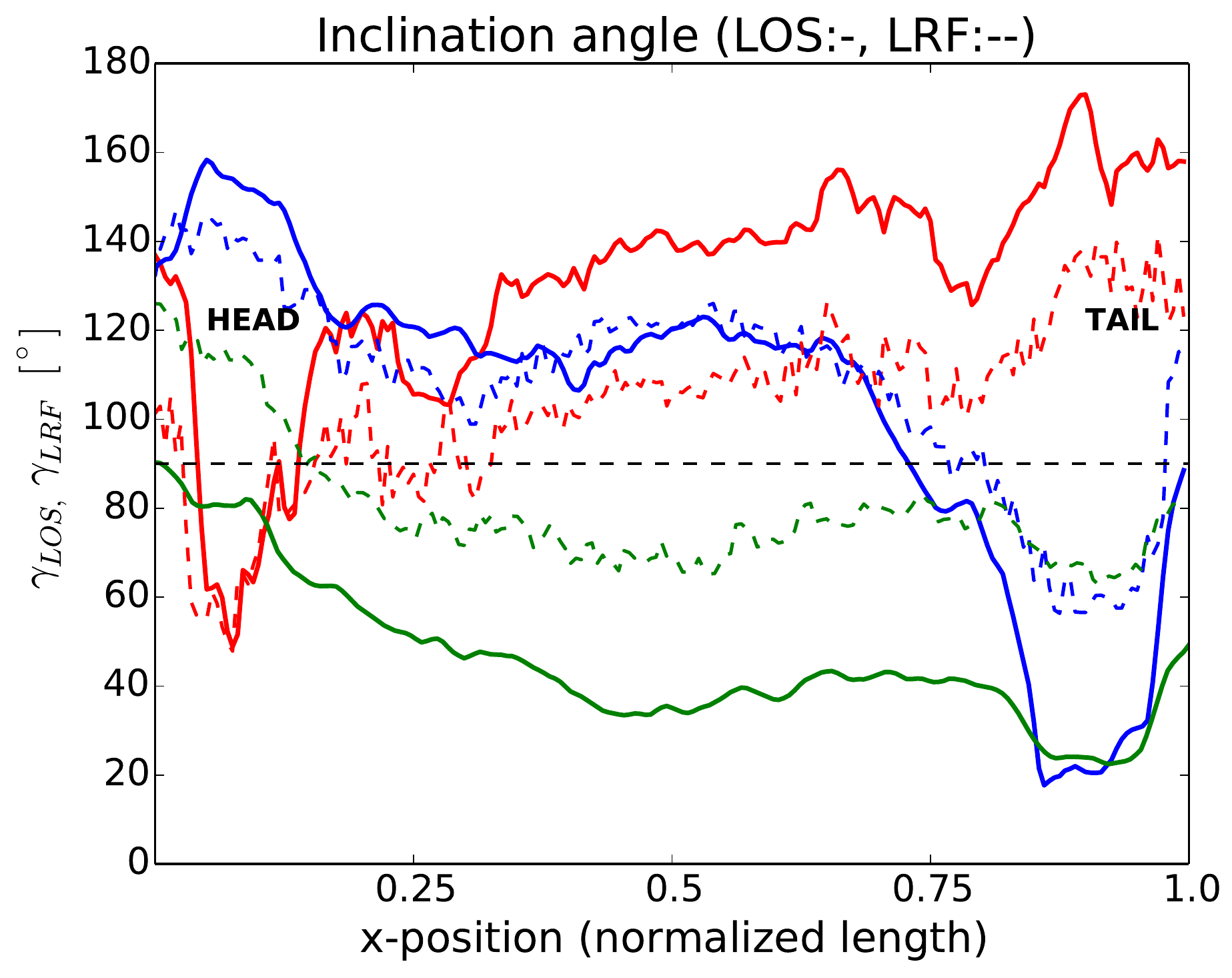}
 }

    \caption{Thermal, magnetic and velocity profiles along the central axes of average filaments at $\log (\tau) = 0$. Clockwise: temperature $T$, magnetic field strength $B$, field inclination in the line-of sight $\gamma_{LOS}$ (solid lines) and in the local reference frame $\gamma_{LRF}$ (dashed lines), and line-of-sight velocity $v_{LOS}$. The profiles correspond to the longitudinal cuts denoted by horizontal dashed lines in Figure \ref{fig:12}:  center-side NEF (blue), limb-side NEF (green) and CEF (red). Note that the "natural"  x-position of the average filament from the CEF case has been reversed so that we now refer to its outermost footpoint (closest to the quiet Sun) as its head and  to the innermost one (closest to the umbra) as its tail.} 
        \label{fig:13}
   \end{figure*}

The LOS velocity shows strong blueshifts ($v_{LOS}\sim-5$ km s$^{-1}$) concentrated in the head of the center-side filament, and weaker blueshifts ($v_{LOS}\sim-2$ km s$^{-1}$) that continue
in a thin band along the central axis of the filament. In contrast, large redshifts ($v_{LOS}>5$ km s$^{-1}$) appear concentrated in the tail. This can also be clearly seen in
Figure \ref{fig:13}, where the thermal, the magnetic and the velocity profiles along the central axes of the three average filaments are plotted  (the corresponding cuts are indicated by horizontal dashed lines in Figure \ref{fig:12}).

From Figures \ref{fig:12} and \ref{fig:13} one can find various similarities between the center-side and the limb-side averaged NEF filaments: the temperature at the head of the limb-side NEF filament is roughly the same as in the center-side NEF case, and it drops from the head towards the tail, while the magnetic field suffers a notable strengthening near  the tail. 
   As in the center-side NEF filament, the limb-side NEF filament shows a slight temperature increase at the tail with respect to the body, assuming temperatures of up to $\sim5800$ K. The field strength reaches values up to $\sim2.2$ kG in the tail.

 Figure \ref{fig:12}  also shows that the flow is mainly concentrated along the body of the limb-side NEF filament, reaching velocities $v_{LOS}>6$ km s$^{-1}$ in the tail. 
Note, however, that in this case we do not observe flows in the heads, likely because the orientation of the magnetic field is almost perpendicular to the LOS direction in that region. Assuming that the flow and the field have the same direction, no flows could then be observed along the LOS.

In addition, unlike in the center-side NEF case, the LOS inclination does not show the opposite field polarity between the head and the tail of the limb-side filament. Instead, we see in the head $\gamma_{LOS} \sim 80^{\circ}$,  $\gamma_{LOS} \sim 40^{\circ}-50^{\circ}$ along the body, and  $\gamma_{LOS} \sim 20^{\circ}$ in the tail. Nonetheless, once the field inclination is transformed into the LRF, the opposite field polarities between the filament endpoints and a roughly horizontal field along the body are unveiled: $\gamma_{LRF} \sim 120^{\circ}$ in the head (slightly downward pointing field), $\gamma_{LRF} \sim 80^{\circ}$ along the body (almost horizontal field) and $\gamma_{LRF} \sim 60^{\circ}$ in the tail (slightly upward pointing field).

\begin{figure}
   \centering
\resizebox{\hsize}{!}{   \includegraphics[width=\hsize]{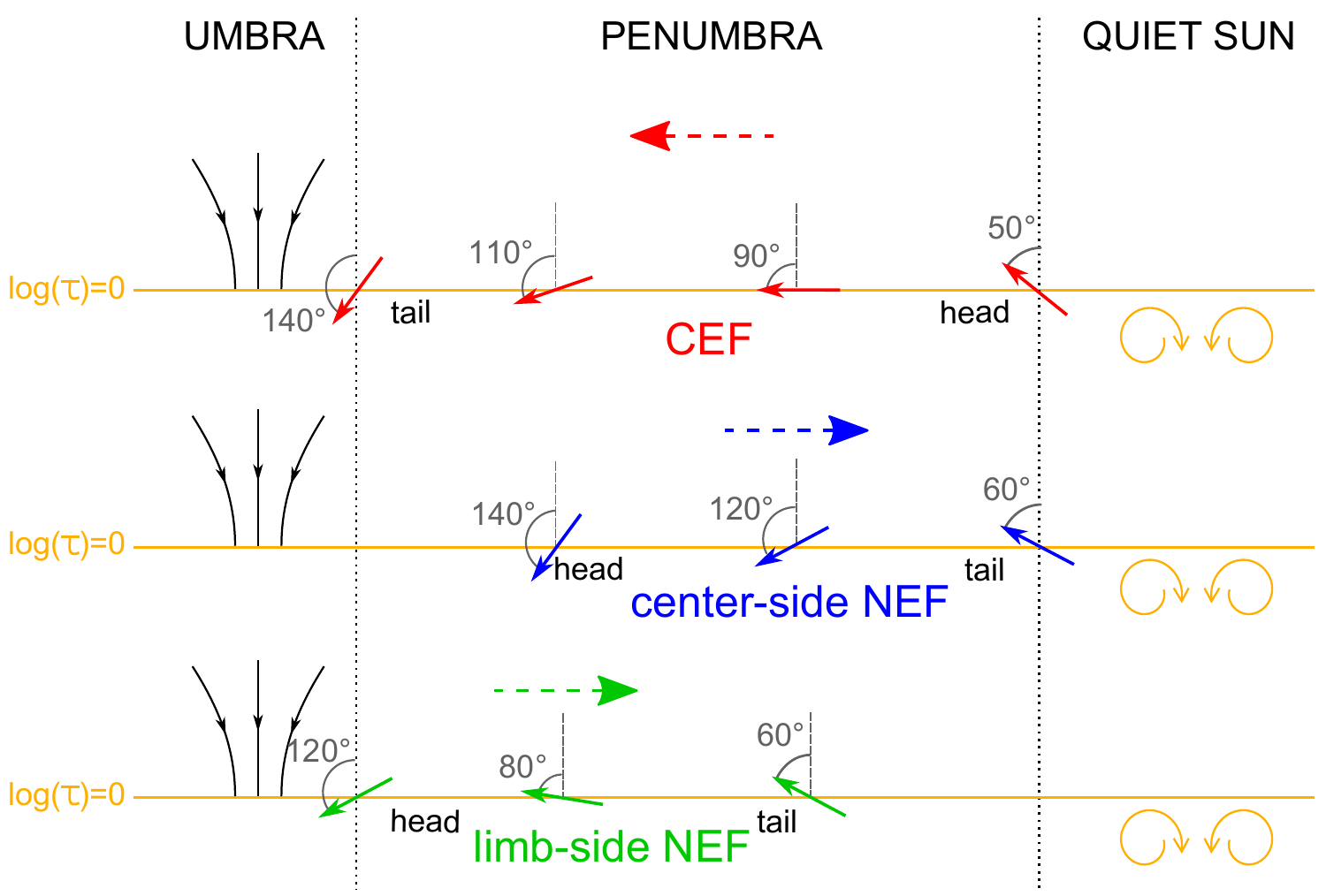}}
      \caption{Possible magnetic field configuration of the three average filaments from different sectors of the penumbra, according to $\gamma_{LRF}$ at $\log(\tau)=0$. Solid arrows represent the magnetic field inclination in the CEF average filament (red), center-side NEF (blue) and  the limb-side NEF (green) average filaments. The dashed horizontal arrows indicate the radial flow direction.      }
         \label{fig:extra0}
   \end{figure}

The slightly different curvatures of the field along the center-side and limb-side NEF filaments, as seen in the LRF (see sketch in Figure \ref{fig:extra0}), cannot be attributed to their different radial locations within the penumbra (outer and inner filaments, respectively) since, according to the findings of \citet{Tiwari2013}, the field inclination is remarkably independent of the location of the filament, displaying only modest differences between filaments at different radial distances.

There are multiple possibilities for the different curvatures of the center-side and limb-side NEF filaments. One possibility is that we see somewhat different heights in the center-side and limb-side penumbra, as the ray passes through different atmospheric structures so that a given optical depth is reached at different heights in the two geometries.  There are also biases introduced by the fact that Stokes $V$ is typically significantly stronger than Stokes $Q$ or $U$, so that a magnetic field directed (anti-) parallel to the LOS gives a larger contribution than one that is directed perpendicular to the LOS. If there is a mixture of unresolved fields with different strengths and inclinations (or if SPINOR does not remove all the stray light), then this introduces a difference in the field strength (compare $B$ in the the heads and tails of the two NEF filaments in Figures \ref{fig:12} and \ref{fig:13}) and also in the inclination. Finally, the penumbra may be 
  intrinsically asymmetric, with the body of the limb-side filaments formed by almost horizontal fields that are  pointing inwards and slightly upwards, and the body of the center-side NEF filaments formed by also inward directed almost horizontal fields but pointing slightly downwards.  

\begin{figure*}
   \resizebox{\hsize}{!}
            {\includegraphics[width=0.5\textwidth]{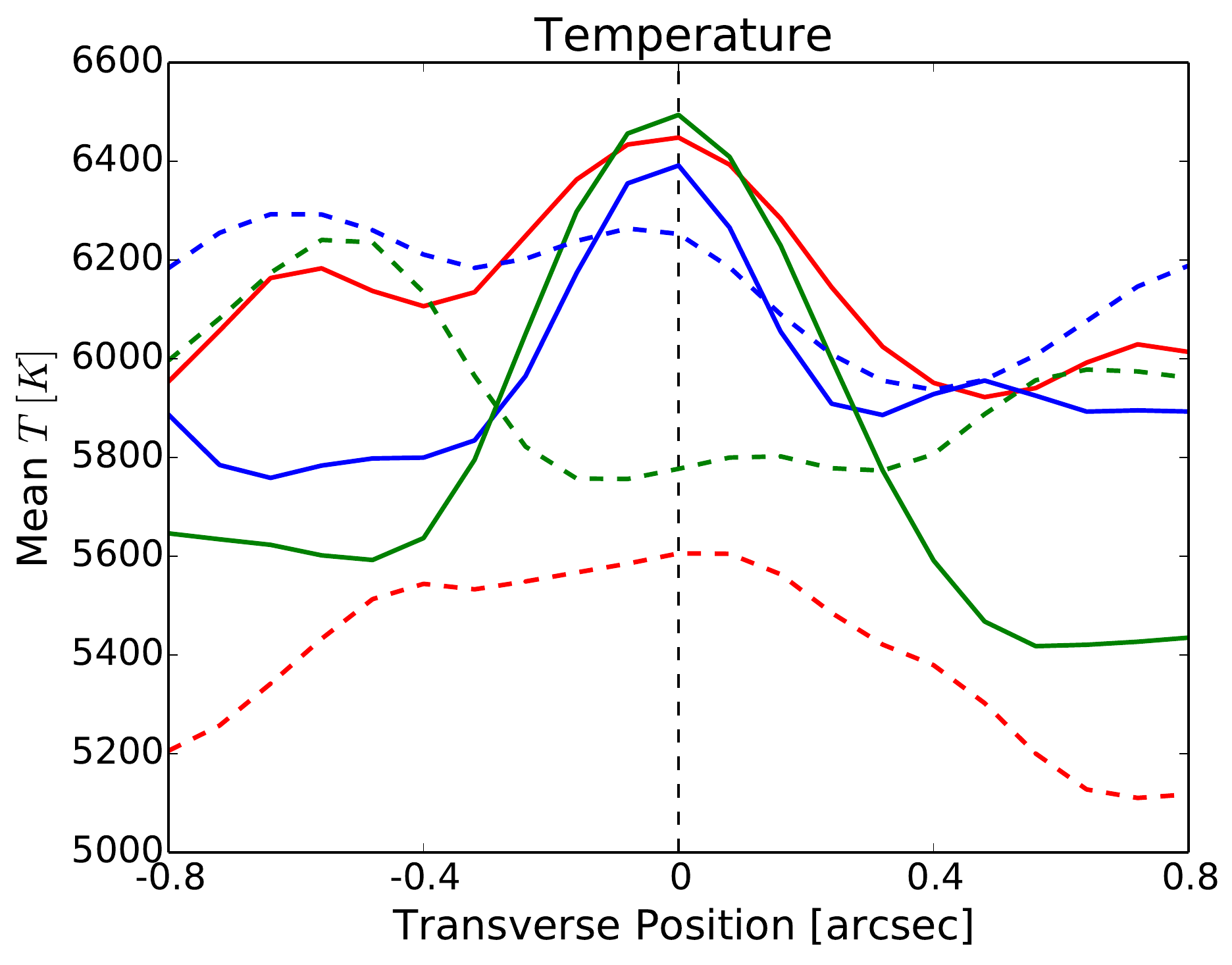}
 \includegraphics[width=0.5\textwidth]{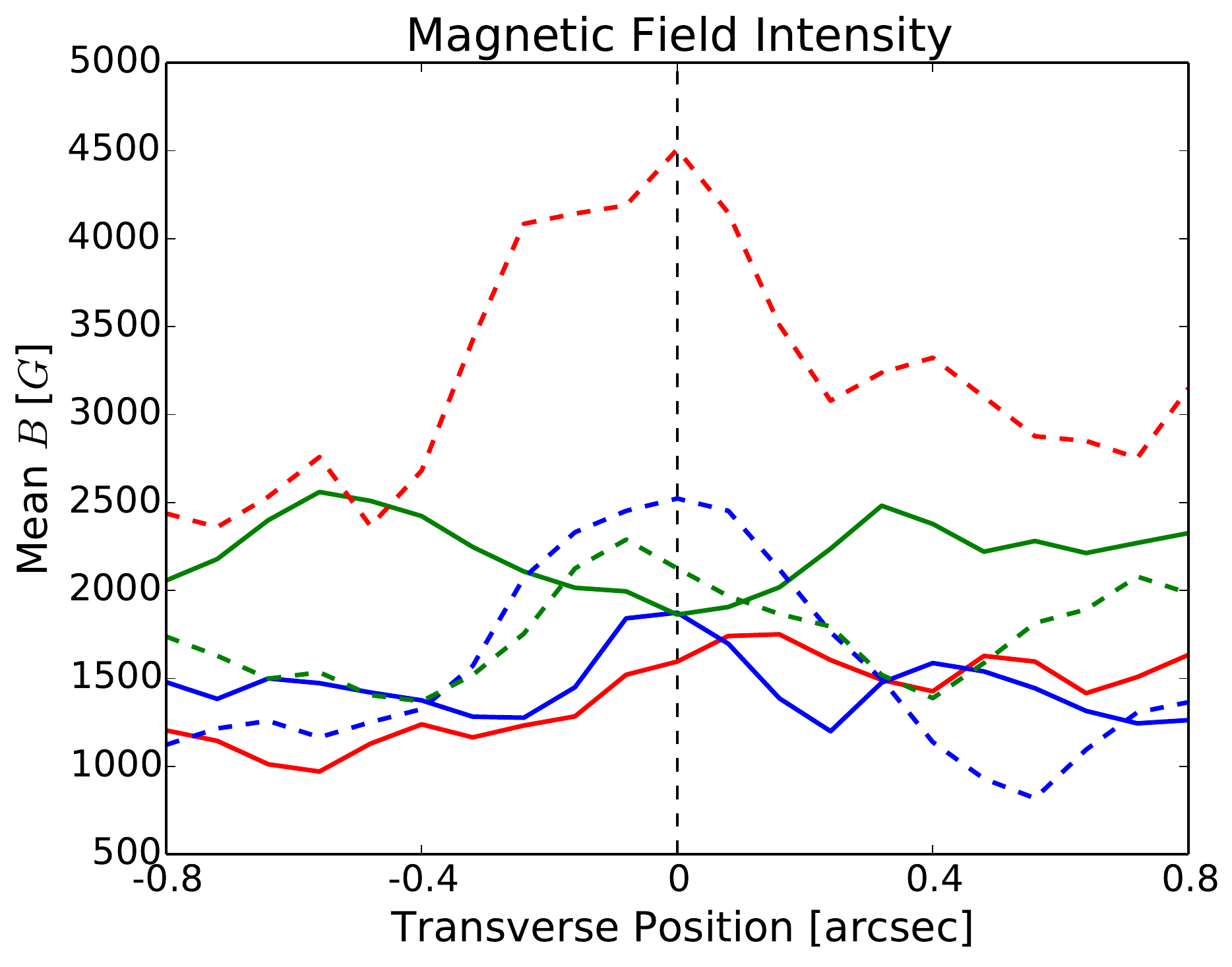}
}

\resizebox{\hsize}{!}
            {\includegraphics[width=0.5\textwidth]{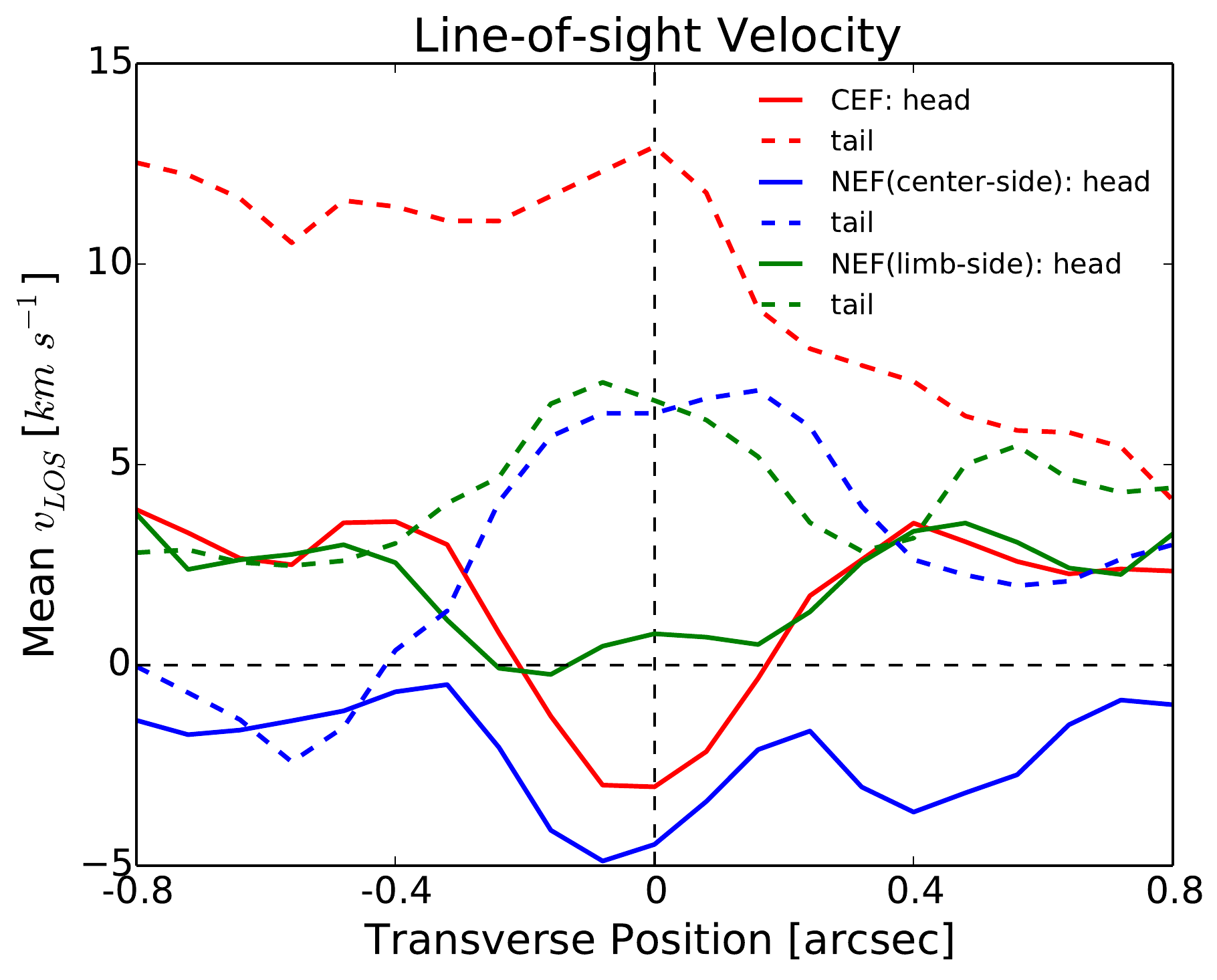}
\includegraphics[width=0.5\textwidth]{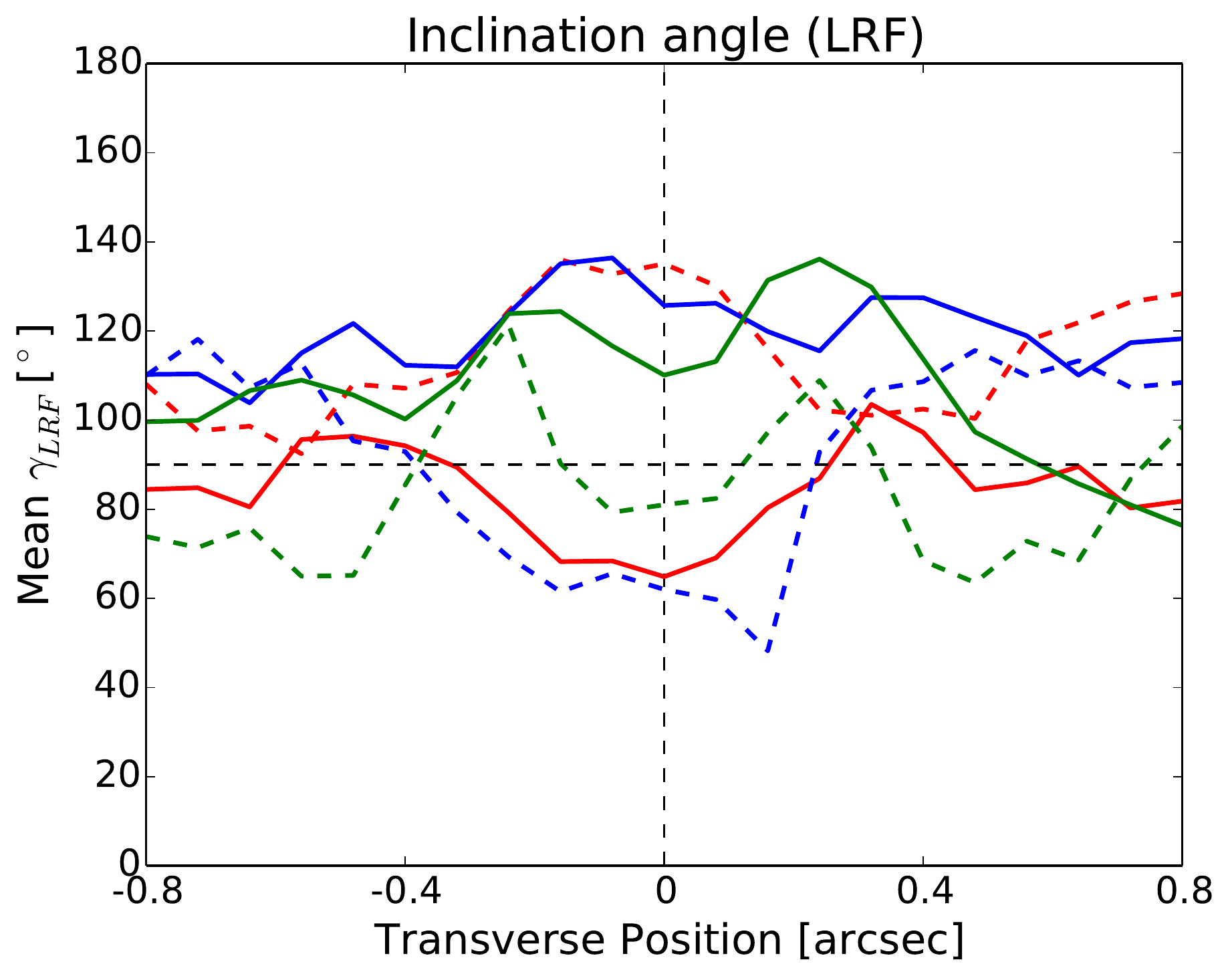}
 }

    \caption{Variation of  $T$, $B$, $v_{LOS}$ and $\gamma_{LRF}$ along the transversal cuts marked by vertical dashed lines in  Figure \ref{fig:12} at $\log (\tau) =
0$: heads (solid lines) and tails (dashed lines). The color designation is the same as in Figure \ref{fig:13}. Black dashed vertical lines represent the central axis of the averaged filaments.
              }
        \label{fig:14}
   \end{figure*}

As expected, there are some remarkable differences seen in the surroundings of the averaged filaments from the center-side and limb-side NEF regions since they are  the result of averaging groups of outer and inner filaments, respectively, and the physical parameters partly change quite considerably from the inner to the outer penumbra. 
Such differences are consistent with the results of  \citet{Tiwari2013} for the surrounding environment of filaments located in the inner and outer penumbra.
Even though the temperature in the surroundings of the filaments increases gradually with radial distance in both cases (i. e. when moving from umbra to the quiet Sun), the environment of the filament from the center-side NEF penumbra (outer filaments)  is significantly hotter than in the limb-side NEF case (inner filaments). 
As argued by \citet{Tiwari2013}, the differences in the properties of the surroundings are  due to the variation of the spines with radial distances. Such differences appear to have an important effect on the downstream temperature structure of the filaments (tail of filament in center-side NEF region is hotter than the tail in the limb-side case). 
This can be clearly seen in both, Figures \ref{fig:13} and \ref{fig:14}. In the latter, various physical parameters are plotted along 2 transversal cuts near the head (solid lines) and the tail (dashed lines), indicated by vertical dashed lines in Figure \ref{fig:12}.

Figure  \ref{fig:14} also shows that the magnetic field  surrounding the head of the limb-side NEF filament (inner filaments)  is significantly stronger than inside the head of the filament. 
In contrast, the magnetic field strength surrounding the head of the center-side NEF  filament (outer filaments) is slightly weaker than the field strength inside the  head. 
However, as clearly revealed by Figures \ref{fig:13} and \ref{fig:14}, the field strength in the filament itself seems to be very similar for both, the center-side and limb-side NEF regions, independently of their different radial locations. 
Moreover, the field strength is $\geq 1$ kG everywhere inside both NEF filaments, and slightly larger than $2$ kG in their tails.

\

\subsubsection{CEF filaments}
Figure \ref{fig:12} also shows that the endpoints of the averaged filament from the CEF region are exchanged compared to the NEF case: the filament endpoint closer to the umbra displays properties similar to the tail of the NEF filaments but being even more extreme, with a local  large strengthening of the magnetic field ($\sim 4.5$ kG) and large redshifts that correspond to supersonic line-of-sight velocities ($v_{LOS}>9$ km s$^{-1}$); while the outer endpoint behaves as a head, being much hotter than its surroundings and harboring concentrated blueshifts.
 This clear spatial anti-correlation strongly suggests that the filaments in the CEF region are "reversed filaments", with their heads located at their outermost endpoints and their tails  at their umbral directed endpoints.

According to Figure \ref{fig:12}, the head of the averaged filament from the CEF region is followed by  a thin band along the center of the filament where relatively low and positive values of $v_{LOS}$ are seen (note that  the head lies close to the outer boundary of the spot and the filament extends towards the umbra). 
The LOS velocity gradually increases from head to tail along the central axis of the filament, from $\sim 1$ to $\sim 5$ km s$^{-1}$, which represents an inflow towards the umbra (counter Evershed flow) roughly along horizontal magnetic field lines ($\gamma_{LRF}\sim 90-110^{\circ}$).

We interpret the observation of low $v_{LOS}$ values just after the head of the filament to be a result of the viewing geometry. In fact, the LOS velocity shows stronger redshifts around the head and in the lateral edges immediately after the head than inside the body of the filament  (this is clearly visible mainly between 0.8  and 0.9 of the normalized x-position of Figure \ref{fig:12}). Those lateral redshifts are likely produced by lateral downflows occurring at the edges of the filament where the field is pointing downwards ($\gamma_{LRF}\sim 140^{\circ}$). The dominance of the lateral downflows on the velocity transversal profile of that region can also be explained by the viewing geometry:
 there, the field
 is more aligned to the LOS direction ($\gamma_{LOS}\sim 150^{\circ}$) than  in those regions of weak redshifts along the central axis of the filament (where $\gamma_{LOS}\sim 110^{\circ}$). 

According to $\gamma_{LRF}$,  the CEF would flow almost horizontally right after the head ($\gamma_{LRF}\sim 90^{\circ}$), assuming a coupled bending of the field and the flow immediately after the head.
The tail in this case shows strong, supersonic LOS flows  that we interpret as downflows, which coincide spatially with a large enhancement of the magnetic field strength  and with downward pointing fields ($\gamma_{LRF}\sim 140^{\circ}$).

\subsubsection{Comparison between NEF and CEF filaments}
Plots in Figure \ref{fig:13} disclose a qualitatively similar behavior of the temperature, field strength and line-of-sight velocity along the central axes of the three averaged filaments, from the "heads"  towards the "tails".

In all three cases, there is a systematic decrease of the temperature from the head towards the body of the filaments. Despite a local temperature increase in the tails, the heads are always hotter than the tails. This temperature difference is, $\Delta T$ $\sim$ $800, 600$ and $100$ K for the CEF, limb-side NEF and center-side NEF case, respectively .
Also, in all three cases, the tails display large to very large enhancements in the magnetic field strength, which, compared to the heads, give  $\Delta B$ $\sim$ $3000, 600$ and $300$ G for the CEF, center-side NEF and limb-side NEF case, respectively. 
Note that, because the head of the CEF carrying filaments is lying at the outer penumbra, a larger $B$ value at its tail compared with the head is expected because $B$ is on average larger by $1-1.5$ kG near the umbra than at the outer penumbral boundary \citep{Solanki2003,Tiwari2015}. This explains most, if not all of the difference in   $\Delta B$  between CEF and NEF carrying filaments.
Something also worth noticing is that the bodies of the filaments are not field free gaps, but they are actually magnetized with strengths above 1 kG everywhere in the penumbra, in agreement with the findings of \citet{Tiwari2013}. 

In all three cases, there are concentrated flows in the tails, with $v_{LOS}>5$ km s$^{-1}$. In particular, the tail of the standard filament carrying the CEF displays supersonic velocities ($v_{LOS}>9$ km s$^{-1}$). Such supersonic flows might be contributing to the enhancements in temperature seen at the tails, maybe due to the formation of shocks as suggested by   \citet{Tiwari2013}.

The  LRF inclination angle profiles show opposite field polarities  between heads and tails in all three average filaments. 
To help the reader  visualize the magnetic configuration of the average filaments at $\log(\tau)=0$, in Figure \ref{fig:extra0} we have sketched the magnetic field inclination of the three average filaments, according to the information provided by $\gamma_{LRF}$ in the deepest visible layer: 
  the head of the center-side and limb-side NEF filaments have both the same polarity as the umbra ($\gamma_{LRF}\sim140^{\circ}$ and $120^{\circ}$, respectively), while the CEF filament head has the opposite polarity ($\gamma_{LRF}\sim50^{\circ}$). 
Note that the magnetic field displays roughly 
the same general magnetic polarity in both the CEF and 
center-side NEF parts of the penumbra when looking 
at both from the umbra outwards. The anti-correlation between their $\gamma_{LRF}$ profiles at the heads and
 the tails of NEF and CEF, seen in Figures \ref{fig:13} and \ref{fig:14}, is a result 
of plotting the filaments such that the heads (i. e. upflows) are 
together.
The different radial position within the penumbra of the center-side and limb-side NEF filaments (outer and inner filaments, respectively) as well as the different radial extent of the CEF filaments, have been sketched in Figure \ref{fig:extra0}.

From the plots in Figure \ref{fig:14}, we can again find considerable similarities in the behavior of the two endpoints of the averaged filaments in different sectors of the penumbra, with the biggest differences being observed  between their surroundings.
 The plasma surrounding the tail of the CEF averaged filament is substantially cooler than in the NEF cases. This is not surprising since most of the filaments in the CEF region are seen to penetrate  into the umbra (see Figure \ref{fig:2}) and consequently the tails are located in a rather cool environment (which is even cooler than the media surrounding the heads of the inner filaments in the limb-side NEF region). 

In summary, we could say that the flows (both, the NEF and the CEF) all start as upflows in the bright and hot head of the filament containing them, where the field is more vertical. The material is then carried radially outwards/inwards along the NEF/CEF filament's axis. 
Along the way, the gas cools and finally sinks again at the tails, where the gas is cooler than in the heads and  the field is stronger.

\begin{figure*}
    \centering
    \includegraphics[width=\textwidth]{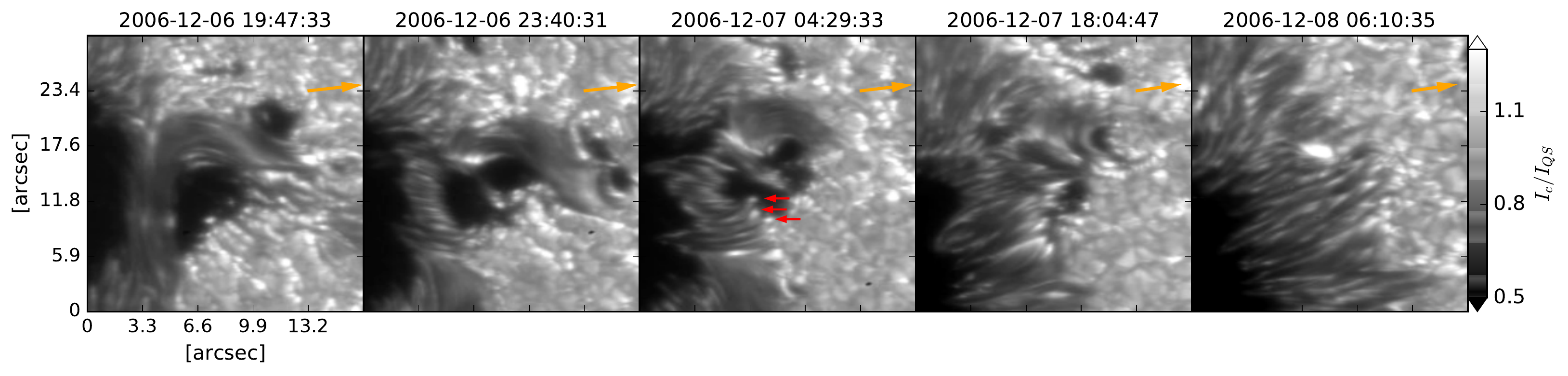} 
    \caption{Hinode G-band images showing the evolution of the anomalous penumbra in AR 10930, at five selected stages. Yellow arrows point towards the disk center. The red arrows in the third frame show the location of penumbral bright grains. 
}\label{fig:extra}
\end{figure*}

\section{Discussion}

In the center-side penumbra, one would normally expect to observe blue-shifts or plasma flows directed towards the observer due to the well-known photospheric Evershed effect \citep{Evershed1909},  a radial  and almost horizontal outward-directed flow of material. However, the observed redshifts in a sector of the center-side penumbra of the main spot of AR NOAA10930
caused by plasma motions pointing away from the observer, indicate either inward  directed motions (towards the umbra) or  downflows in the center-side penumbra. 
Under the assumption of flows being 
aligned with the magnetic field, such a red-shifted region in the center-side penumbra represents a counter-Evershed flow.
We  discard the possibility that these redshifts could be produced by a nearly vertical downflow of plasma only, due to the finding of an associated radial penumbral fine structure (i. e. filamentary fine structure) and further similarities with the normal Evershed outflow.  Nonetheless, the observation of strong redshifts in regions with more vertical fields at the edges of the filaments carrying the CEF suggests the existence of lateral downflows channels.

The very large line shifts and line splittings observed in the anomalous penumbral region, and the extensive area that the CEF spans, make this sunspot a rather unique one.
 The previously reported observations of CEFs  by, e. g., \citet{Kleint2013,Louis2014} have been mainly restricted to
 penumbral filaments of sunspots leading to flares, or have been found in forming penumbrae \citep{Schlichenmaier2012,Romano2014,Murabito2016}.

\citet{Kleint2013} propose two possible models to explain the observation of CEFs in the photosphere. The first model is  
the \textit{umbral filament sheet model}, in which the CEF would occur along a topological feature in form of a sheet that magnetically connects spatially separated regions (e. g., the umbra with a network element well outside the spot), the CEF is then described as a siphon flow produced  by the pressure difference between the umbral base of the sheet and the magnetically connected network element. The second model proposed by  \citet{Kleint2013} is the \textit{massive umbral filament model}, in which a thick flux tube with higher density than the penumbra carries the common chromospheric inverse EF, with the NEF hidden below the flux tube.

In our study, we have not  analyzed the dynamics of the anomalous part of the penumbra in the higher layers of the atmosphere. We have concentrated mainly on the deepest visible layer ($\log(\tau)=0$), where the sources of the CEF could be identified
 within the penumbra itself (at the outer border).  
On the one hand, the fact that we see a CEF at the deepest observable layers in an otherwise nearly normal photospheric penumbra, is not compatible with  the  \textit{massive umbral filament model} of \citet{Kleint2013}, which basically describes the CEF as a chromospheric flow. On the other hand, the fact that the CEF begins and ends within the penumbra itself, is conflicting with the \textit{umbral filament sheet model}, which requires that the flow is driven from a network element.
Also, we do not observe turbulent motions in the boundary between the CEF and the NEF (perhaps due to the spatial resolution), as would be expected in the \textit{umbral filament sheet model}.
Instead, we observe that the CEF is concentrated along penumbral channels, whose magnetic and thermal structures strongly suggest that they are "inverted normal filaments" with their heads/flow-sources located at the outer penumbral boundary and the tails/flow-sinks located in the inner penumbral boundary.

Our analysis allows for two interpretations of the driver of the CEF (and the NEF). One possible qualitative picture of the CEF (and of the NEF) emerging from our analysis is that of a siphon flow driven by a gas pressure gradient due to different magnetic field strengths at the two endpoints of the flux tubes forming the penumbral filaments in the anomalous region, in accordance with the model of  \citet{Meyer1968} for explaining the driving forces of the NEF.

We found a  field strength gradient of $\sim 3000$ G between the tail and the head of the averaged filament carrying the CEF ($\sim 300-600$ G in the averaged filaments carrying the NEF). This may contribute to the acceleration of the flows since an enhanced field in the tails implies a larger gas pressure in the head compared to the tail. However, to validate this siphon flow scenario, we need to know if a pressure gradient exists between the endpoints of the filaments at the same gravitational potential, i. e., at constant geometrical height. This is unfortunately not possible to know since the present observations provide physical information of constant optical depth layers only.

According to \citet{vannoort2013}, the strongest fields in penumbrae are usually found at the
ends of complex filaments, particularly those with multiple
heads that merge to form a single tail. Those tails show a polarity opposite to that of the sunspot umbra and contain supersonic downflows ($\geq 9$ km s$^{-1}$). 
\citet{vannoort2013} argue that the strong magnetic fields are probably the result of intensification of magnetic field by the  collapse of magnetized flux concentrations \cite[e.g.][]{Parker1978}.  In addition,  the optical depth unity surface might be strongly depressed at the tails of the filaments, exposing stronger fields from a deeper geometrical height.
In our study, we have only analyzed simple filaments (with a single head and a single tail) carrying the NEF and the CEF, respectively. The tails of these filaments also contain enhanced field strengths ($B\sim 2-2.5$ kG on average for the tails of NEF carrying filaments and $B\sim4.5$ kG on average for the tails of CEF carrying filaments), and are co-located with fast downflows (within the subsonic regime for the NEF case, $v_{LOS}\sim7$ km s$^{-1}$ on average; and with supersonic speeds in the CEF case, $v_{LOS}>9$ km s$^{-1}$). 
The mechanisms considered by \citet{vannoort2013} are also a possible explanation for the relation between the large downflow velocities and the enhanced magnetic field strengths in our observations.
Likewise, it is possible that the enhanced field strengths found in the tails of the filaments carrying both the CEF and the NEF, correspond to regions below the average geometrical height of the penumbra, as proposed by  \citet{vannoort2013} and  \citet{Tiwari2013}.
  This could also explain the temperature enhancement found in the tails of the filaments, since we might be seeing deeper and hotter layers in the tails 
than in the other parts of the penumbra. The supersonic downflows observed in the tails of the filaments carrying the CEF might contribute to the increase in temperature due to the formation of shocks.

In particular, we do not discard the possibility that the very large magnetic field values returned by the SPINOR 2D inversions ($B>7$ kG) 
are real and they are observed in the penumbra due to an unusually depressed optical depth surface formed as the consequence of very low densities in the downflowing part of the anomalous penumbra harboring the CEF. 
However, given that most of the pixels where SPINOR 2D returns  $B>7$ kG are located at or close to the umbral/penumbral boundary of the CEF region (see yellow markers in Figure \ref{fig:3b*}) and contain very complex Stokes profiles (e. g. they exhibit a large wavelength separation, large asymmetries and multi-lobed Stokes $V$ profiles),
it is also possible that those profiles are produced by multiple (unresolved) atmospheric components with large differences in their Doppler velocity. 
One of the components could be associated with the umbral magnetic field in the sunspot (where the medium is nearly at rest) and the second one with the filamentary penumbra (strongly red-shifted component). Very large Doppler shifts as well as extremely strong magnetic fields could explain the large wavelength separation observed at these peculiar pixels in the umbral/penumbral boundary.  However, in order to get insight on the nature of these complex profiles it is necessary to perform a detailed analysis using e. g. some classical diagnostic methods and different inversion techniques considering different model atmospheres to see which one gives the most reliable results.
This will be the topic of a future study.

The other possible driver compatible with our results is the thermal gradient. The  systematic temperature decrease from the heads to the tails observed in all three averaged filaments (limb-side NEF, center-side NEF and CEF) is compatible with the convective driver scenario, as proposed by \citet{Scharmer2006,Spruit2006}, but modified by the presence of a magnetic field since we observe field strengths $B>1$ $kG$ in the body of the filaments, similar to the findings of  \citet{Tiwari2013} in filaments carrying a NEF. 
In this scenario, the upflowing hot gas reaches the solar surface due to the convective instability. There, the gas decelerates and builds up excess pressure. Due to the generally radial and horizontal magnetic field direction, the gas flows to a large extent radially along the body of the filaments. Along the way, the gas cools down  and it eventually sinks in the tails. This is in qualitative agreement with the results of the simulations of \citet{Rempel2009b}.
Independently of their  opposite horizontal flow direction, in both cases (NEF and CEF), the relationship between the direction of flow and the temperature,
with upflowing material being hotter than the downflowing
material, could provide support  for the presence of overturning
convection along the penumbral filaments. 
Furthermore,  since we might be seeing higher layers in the heads than in
the tails, the temperature difference at an equal geometrical
height between heads and tails should be considerably larger than the observed \citep{Bruls1999}. 
However, the relevant physical parameters need to be known on a geometrical scale  to confirm this scenario.

The various opposite field polarity patches that are observed outside but in the vicinity of the anomalous part of the penumbra, might be related with the initiation of the CEF, since this is apparently the only aspect that distinguishes (in the surroundings) the part of the penumbra with CEF from the other parts of the penumbra considered as normal. The lack of SP scans prevents us from observing the exact time at which the CEF is initiated in the anomalous part of the penumbra. However, Hinode observations of the G-band and narrow-band filtergrams show the presence of an adjacent pore with opposite magnetic polarity to that of the umbra of the main sunspot prior to the analyzed SP scan. This pore was located in the region where the opposite polarity patches are observed in the present SP scan (green contours in Figure \ref{fig:3a*}), which might be remnants of the pore. 
A quick look into the temporal evolution of the pore in the G-band images (see Figure \ref{fig:extra}) suggests that the pore develops a penumbra-like connection with the main sunspot (visible in the second frame). The CEF carrying part of the penumbra developed out of this initial connection.
The third frame shows that the penumbral filaments in the anomalous region grow while the area of the pore decreases. Note that the location of the penumbral bright grains (red arrows in third frame) suggests that the filaments originate in the pore and extend outwards, towards the umbra of the main sunspot. Consequently, if a NEF is carried along those filaments (from the pore outwards), that would mean an inflow towards the umbra of the main sunspot in the AR, i. e., a CEF. The fourth frame shows that the area of the pore continues decreasing while the penumbral filaments grow, and finally, in the last frame (which roughly corresponds to the time of the analyzed SP scan) the pore has disappeared and the penumbral filaments in the anomalous region seem to have been "adopted" by the main sunspot of the AR while carrying a CEF.

\citet{Jurcak2017} also reported the evolution of a penumbra at the boundary of a small pore, in which the penumbra seemed to colonize the pore area leading to its extinction. They found that the maximum value of the vertical component of the magnetic field $B_{ver}$ in the pore was around 1.4 kG and  argued that a stable umbra-penumbra boundary could not be formed in that case because the pore did not fulfill  the canonical critical value of $B_{ver}=1.8$ kG, empirically found by \citet{Jurcak2011} and \citet{Jurcak2015} to be a crucial value for the formation of a stable umbra-penumbra boundary in a magneto-convective context.
In our case, we do not have enough information on the magnetic field configuration in the pore during the formation process of the anomalous penumbra due to the lack of SP scans during those stages. Nonetheless,
 its evolution on the G-band images looks in some aspects similar to the  case studied in \citet{Jurcak2017}: the filaments seem to grow at the expense of the adjacent pore. However, an important difference is that in the  case studied by \citet{Jurcak2017}, the penumbra ends up as an orphan penumbra once the pore has disappeared, i. e. the filaments are not connected to any umbral region, while the anomalous penumbra in our study is continually attached to the umbra of the main sunspot.

The evolution of the anomalous penumbra in AR 10930 is  associated with high chromospheric activity. The study of the associated chromospheric and coronal dynamics, from the time of the AR's  first appearance over the solar west limb on 2006-12-06 until the disappearance of the CEF on 2006-12-09 when the center-side penumbra shows the NEF only,
might provide us with important information on how the anomalous penumbra was formed and  the CEF maintained during a couple of days before reversing into a NEF. This will be the topic of a future work.

\section{Conclusion}
In this paper, we have reported the observation of a red-shifted region in the center-side penumbra of the main sunspot of the AR NOAA 10930 on 2006-12-08 06:11:14 UT, at photospheric heights.
This is, to our knowledge, a unique observation of a counter Evershed flow (CEF) at the photospheric heights covering a sizable part 
of the penumbra of a mature sunspot.

  By using the SPINOR 2D code to invert spectropolarimetric data of the sunspot from the Hinode SOT/SP instrument,  we  investigated the characteristics of the CEF in the photosphere and have compared them with the physical properties of the normal Evershed flow (NEF) in the same sunspot.

The results found here are consistent with both scenarios, namely that
 the temperature gradient or the magnetic field gradient is the main driver of both, 
  the normal Evershed flow and of the anomalous or counter Evershed flow. This implies that we cannot distinguish between 
 the convective driver and the siphon flow scenarios.
 However, this result suffers from the shortcomings of present observations and analysis techniques: the inability to peer below the solar surface and the lack of knowledge of the true geometrical height scale.

A comparison of our results with recent high-resolution sunspot simulations that display a NEF in the penumbra and a number of transient regions with a CEF at photospheric heights \cite[see][]{Rempel2015} will provide us with additional tools to determine the dominant forces driving the flows.
An important advantage of this is that we will be able to study the dynamics of the flows at constant geometrical heights and to look into the vertical gradients at sub-photospheric depths, which are not accessible to observations.

Finally, studying in detail the history of the sunspot and the associated chromospheric activity can help us  understand the formation process of the anomalous penumbra, the initiation of the CEF and its change of direction into a NEF.  This will be the topic of a future study.

\begin{acknowledgements}
We thank N. Bello Gonz\'alez and R. Schlichenmaier for fruitful discussion and suggestions about the possible origin of CEF. We also thank the referee of this paper for providing insightful comments and directions for additional work which has substantially improved the presentation.
This work was carried out in the frame of the International Max Planck Research School (IMPRS) for Solar System Science at the Max Planck Institute for Solar System Research (MPS). It is supported by the Max Planck Society and by BECAS CONACyT AL EXTRANJERO 2014.
This work was partly supported by the BK21 plus program through the National Research Foundation (NRF) funded by the Ministry of Education of Korea.
\textit{Hinode} is a Japanese mission developed and
launched by ISAS/JAXA, with NAOJ as a domestic partner and
NASA and STFC (UK) as international partners. It is operated
by these agencies in co-operation with ESA and NSC (Norway). 
\end{acknowledgements}

\bibliographystyle{aa} 

\bibliography{bib_monita2}{}

\end{document}